\documentclass[apj,numberedappendix]{emulateapj}
\usepackage{graphicx, bm, color, amsmath, amssymb, xfrac}
\usepackage{enumerate, enumitem}
\usepackage{multirow}
\usepackage[usenames,dvipsnames]{xcolor}
\usepackage{hyperref}
\hypersetup{
    colorlinks = true,
    citecolor = {MidnightBlue},
    linkcolor = {BrickRed},
    urlcolor = {BrickRed}
}
\usepackage{times} 

\def\corr{}

\newcommand{\be}{\begin{equation}}
\newcommand{\ee}{\end{equation}}
\newcommand{\bea}{\begin{eqnarray}}
\newcommand{\eea}{\end{eqnarray}}

\begin{document}

\title{Extending cosmological tests of General Relativity \\
with the Square Kilometre Array}
\shorttitle{Extending cosmological tests of GR with the SKA}

\author{Philip Bull}
\affil{Institute of Theoretical Astrophysics, University of Oslo, P.O.\ Box 1029 Blindern, N-0315 Oslo, Norway}
\email{p.j.bull@astro.uio.no}


\begin{abstract}
Tests of general relativity (GR) are still in their infancy on cosmological scales, but forthcoming experiments promise to greatly improve their precision over a wide range of distance scales and redshifts. One such experiment, the Square Kilometre Array (SKA), will carry out several wide and deep surveys of resolved and unresolved neutral hydrogen (HI) 21cm line-emitting galaxies, mapping a significant fraction of the sky from $0 \le z \lesssim 6$. I present forecasts for the ability of a suite of possible SKA HI surveys to detect deviations from GR by reconstructing the cosmic expansion and growth history. SKA Phase 1 intensity mapping surveys can achieve sub-1\% measurements of $f\sigma_8$ out to $z\approx 1$, with an SKA1-MID Band 2 survey out to $z \lesssim 0.6$ able to surpass contemporary spectroscopic galaxy surveys such as DESI and Euclid in terms of constraints on modified gravity parameters if challenges such as foreground contamination can be tackled effectively. A more futuristic Phase 2 HI survey of $\sim10^9$ spectroscopic galaxy redshifts would be capable of detecting a $\sim 2\%$ modification of the Poisson equation out to $z\approx 2$.
\end{abstract}

\keywords{ cosmology: observations --- large-scale structure of universe --- gravitation}

\section{Introduction} \label{sec:intro}

General Relativity (GR) has been tested to high precision over a comparatively narrow range of scales and redshifts. To date, the strongest constraints come from Solar System and binary pulsar experiments, which represents only a small region in the space of gravitational potentials and spacetime curvature spanned by gravitational phenomena \citep{2008LRR....11....9P, Baker:2014zba}. The detection of deviations from GR would have important ramifications for fundamental physics, so extending tests of gravity to other regimes is of great importance.

This is particularly the case in cosmology, where the discovery of the accelerating expansion of the Universe has highlighted the gaps in our understanding of gravitational physics on the largest scales. As a result, alternative theories of gravity have proliferated in recent years \citep{Clifton:2011jh, 2015PhR...568....1J}, with many of them seeking to replace the need for a cosmological constant or dark energy in order to explain the acceleration. Solar System/pulsar experiments have been valuable in thinning out the extensive list of contenders, as only the theories that are able to almost exactly reduce to GR on these scales can survive the observational constraints. Most viable modified gravity (MG) theories therefore feature `screening mechanisms', which cause deviations from GR to switch off on small scales \citep{2010arXiv1011.5909K, 2012arXiv1211.5237B}, leaving them with significantly different predictions from GR only over cosmological distances.\footnote{Depending on the screening mechanism, deviations from GR can also appear in certain non-cosmological regimes, for example inside evolved stars \citep[e.g.][]{2011ApJ...732...25C} and dwarf galaxies \citep[e.g.][]{2011JCAP...10..032J}.}

Recent searches for observational signatures of MG have primarily focused on understanding how alternative gravitational theories affect several key phenomena: cosmic expansion, the growth of large-scale structure, and light propagation. Distance measurements from the CMB, baryon acoustic oscillations, and Type Ia supernova surveys have been used to reconstruct the expansion history with good precision \citep{2015arXiv150201590P}, but GR+$\Lambda$CDM expansion histories are very often contained within the parameter spaces of MG models, so many of them cannot be definitively distinguished from GR based on the background evolution alone \citep[e.g.][]{2005PhRvD..71d3503C, 2007JPhCS..66a2005N, 2007PhRvL..98l1301K}. More decisive is the linear growth rate, which is increasingly well constrained by redshift-space distortions (RSDs) and other peculiar velocity measurements from large-scale structure surveys \citep{2013MNRAS.429.1514S, 2013MNRAS.436...89R, 2015arXiv150406885J}. Precision tests with large weak lensing surveys have also been performed \citep{2009MNRAS.395..197T, 2013MNRAS.429.2249S}.

MG theories also tend to change the effective strength of gravity on non-linear scales, $k \gtrsim 0.1$ Mpc$^{-1}$, and the modifications can be significant even if linear scales are left relatively unmodified from the GR case \citep{2006PhRvD..74h4007S, 2008PhRvD..77b4048L, 2013JCAP...04..029B, 2014PhRvD..89h4023L}. It is difficult to separate out the contribution of MG from other, less exotic processes, however, due to the difficulty of correctly modelling baryonic effects on small scales, or model confusion with massive neutrinos \citep{2014MNRAS.440...75B, 2014PhRvD..90b3528B}.

Cosmological tests of GR are still in their infancy, and significant increases in precision -- as promised by forthcoming experiments like Euclid, DESI, and LSST -- will be required before many theories can be ruled in or out with confidence. There are few clues as to where one might expect deviations from GR to manifest themselves, and so broadening the reach of the tests in scale and redshift is also necessary -- the ultimate goal being to leave MG theories with `nowhere to hide'. To date, the most stringent cosmological constraints have come from linear scales ($0.01 \lesssim k \lesssim 0.1$ Mpc$^{-1}$), late times (the acceleration era, $z \lesssim 1$), and small/intermediate areas of sky ($\sim\,$few$\,\times 10^3$ deg$^2$), as well as the CMB \citep{Zuntz:2011aq, 2014PhRvL.113a1101A, 2015arXiv150201590P}, but there are many other places one can look. For example, novel gravitational phenomena can arise at late times on scales of order the horizon size, and over wide angular separations \citep{2013PhRvD..87f4026H, Lombriser2013, 2015arXiv150600641B}. Modifications of expansion and growth could also appear at higher redshifts, due to the non-trivial time evolution of (e.g.) an extra scalar field. Adding observations of these as-yet unexplored regimes can only improve our chances of seeing some kind of anomalous gravitational behaviour.

The aim of this paper is to investigate how cosmological tests of GR can be improved and extended to other regimes using a new class of large scale structure survey at radio frequencies. These will use large radio telescope arrays with low-noise wideband receivers \citep{2009astro2010S.219M} to map out the redshift-space matter distribution over a gigantic volume, providing precision data over significantly wider survey areas and redshift ranges than has been possible before. Our focus will be on the Square Kilometre Array (SKA), a planned general-purpose array split over two main sites in South Africa and Australia. The first phase of construction, due to finish around 2023, will consist of two sub-arrays: SKA1-LOW, a low-frequency aperture array operating at $\lesssim 350$ MHz; and SKA1-MID, a conventional mid-frequency array of 130 dishes equipped with low noise receivers covering $\sim\!350$ MHz -- 14 GHz. A second phase, scheduled for completion around 2030, will improve the overall sensitivity by a factor of $\sim 10$.

The SKA will survey large scale structure primarily by detecting the redshifted neutral hydrogen (HI) 21cm emission line from a large number of galaxies out to high redshift. This can be achieved in two ways: by measuring the 21cm line for many individually-detected galaxies (a galaxy redshift survey); or by measuring the large-scale fluctuations of the integrated 21cm intensity from many unresolved galaxies (intensity mapping; IM). The SKA surveys will cover a combined survey volume and redshift range that is significantly larger than that of even Euclid and LSST (albeit with varying, and sometimes lower, sensitivity), and should even begin to probe scales of order the horizon size, $k \sim \mathcal{H}$. Here, we will focus on background observables and linear scales in the post-reionisation Universe, $0 \le z \lesssim 6$; the detectability of horizon-scale modified gravity effects with the SKA will be examined in a forthcoming paper.

This paper is organised as follows. In Sect.~\ref{sec:obs}, we describe a set of large-scale structure observables that can be used to test GR, and outline some parametrisations that can be used to connect them with MG theories. In Sect.~\ref{sec:fisher} we give an overview of the planned SKA HI surveys, and describe our Fisher forecasting methodology. We present forecasts for the various observables in Sect.~\ref{sec:results}, and conclude in Sect.~\ref{sec:discussion}.

\section{Modified gravity observables} \label{sec:obs}

In this section, we discuss some of the generic effects that modifications to GR have on the cosmological background and growth of matter fluctuations in several different regimes. Our primary focus is on linear sub-horizon scales where, for simplicity, we will employ the quasi-static approximation\footnote{Time derivatives of metric perturbations and new degrees of freedom are assumed to be of order $\mathcal{H}$, and subdominant to spatial derivatives \citep[e.g.][]{2013PhRvD..87j4015S}.} and ignore wide-angle corrections \citep[e.g.][]{2013MNRAS.436...89R, 2013PhRvD..87f4026H}.

\subsection{Background expansion history}

Most modified gravity theories exhibit background dynamics that can deviate from the standard $\Lambda$CDM evolution. A simple example is that of models that introduce a new scalar degree of freedom; this is allowed to evolve dynamically, and can therefore have a non-trivial redshift-dependent equation of state \citep{1988PhRvD..37.3406R, 1998PhRvL..80.1582C}. Much depends on the particular structure of the modified theory -- a variety of choices regarding the shape of scalar field potentials, couplings to the matter sector etc. can be made relatively freely, leading to complicated parameter spaces that are often difficult to characterise even for a single theory.

\begin{figure}[t]
\hspace{-2em}\includegraphics[width=1.08\columnwidth]{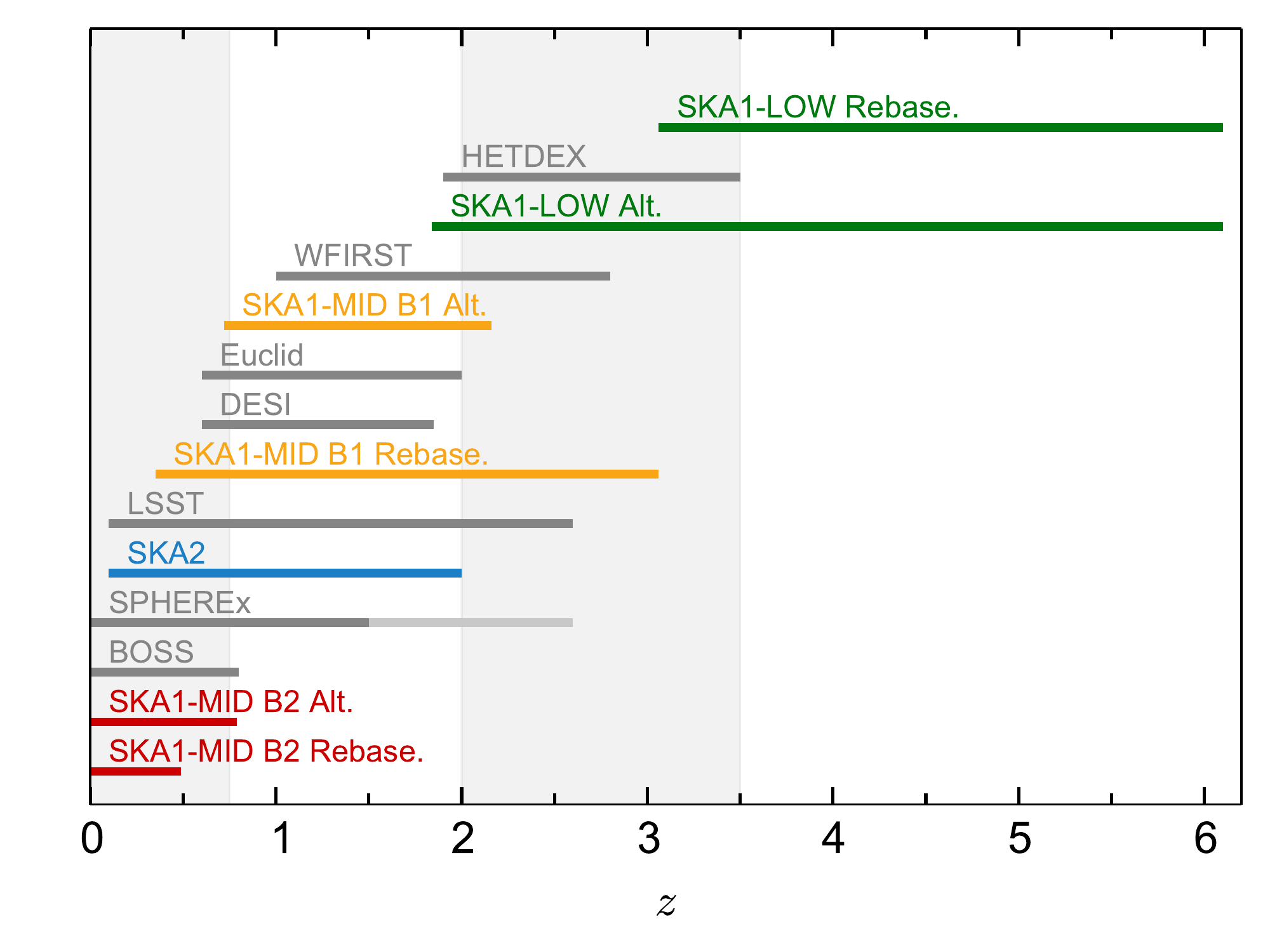}
\caption{Redshift coverage of possible SKA HI intensity mapping and galaxy surveys (colours), compared with other future galaxy surveys (grey). The SPHEREx line shows the redshift range for two galaxy samples.}
\label{fig:surveyz}
\end{figure}

One simple way of modelling the wide variety of possible modifications to the expansion history is by introducing a effective dark energy equation of state parameter, which is commonly parametrised as $w(a) \approx w_0 + w_a (1 - a)$ \citep[CPL;][]{Chevallier:2000qy, 2003PhRvL..90i1301L}. The primary virtue of this parametrisation is its simplicity; this functional form for $w(a)$ is rarely an acceptable fit at all redshifts \citep[e.g.][]{Marsh:2014xoa}, and a variety of potentially important effects (e.g. the sound speed for a scalar field) are neglected. Nevertheless, forecasts for $w_0$ and $w_a$ are a useful way of comparing the relative performance of different survey methods in reconstructing the expansion history, and so we adopt this parametrisation in what follows. Dark energy and modified gravity constraints with more general parametrisations have been considered elsewhere \citep[e.g.][]{2015arXiv150103840Z}.

The equation of state, $w(a)$, can be reconstructed from the angular diameter distance-redshift relation, $D_{\rm A}(z)$, and the expansion rate,
\be
H(a) = H_0\sqrt{\Omega_{M,0} a^{-3} + \Omega_{\rm DE}(a) + \Omega_K a^{-2}},
\ee
where the fractional energy density of dark energy in the CPL parametrisation is
\be \label{eq:omegade}
\Omega_\mathrm{DE}(a) = \Omega_\mathrm{DE, 0}\, \exp\left [3 w_a (a - 1)\right ] a^{-3(1 + w_0 + w_a)}.
\ee
Typically, a number of probes are combined to pin down these quantities over a range of redshifts; we will concentrate on observations of `statistical standard rulers' in the redshift-space matter distribution, combined with a high-redshift constraint from the primary CMB power spectrum. The most robust such standard ruler is the baryon acoustic oscillation (BAO) feature in the matter correlation function. If its true physical scale is known, the observed redshift-space locations of the BAO feature in the radial and transverse directions can be used to constrain two independent combinations of $D_{\rm A}$ and $H$ as a function of redshift, with results that are remarkably robust to systematic effects. Other standardisable rulers can be constructed from the matter distribution -- the overall (`broadband') shape of the power spectrum contains more information than the BAO alone, for instance -- but significantly more careful modelling is required to avoid systematic biases. Optimistically, we will focus on the broadband case here.

\subsection{Linear growth and RSDs} \label{sec:growth}

Gravitational infall is the dominant process in structure formation and so, as one might expect, modifying the theory of gravity can have wide-ranging effects on the cosmic matter distribution. At linear order in perturbations, a relatively general approach to incorporating MG effects is to modify the growth equation (valid on sub-horizon scales),
\be \label{eq:growth}
\ddot{\Delta}_M + \mathcal{H} \dot{\Delta}_M = \frac{3}{2} \mathcal{H}^2 \Omega_M(a) \,\mu(k, a) \Delta_M,
\ee
where $\Delta_M$ is the gauge-invariant matter perturbation and overdots denote differentiation with respect to conformal time (see, e.g., \cite{Baker:2013hia} for a derivation). The function $\mu(k, a)$ has been introduced to parametrise time- and scale-dependent deviations from GR growth. In the GR limit, $\mu \to 1$. We have also defined $\Omega_M(a) = \Omega_{M,0}a^{-3} / E^2(a)$, where $E(a) = H(a)/H_0$ is the dimensionless expansion rate, and $\mathcal{H} = a H$.

The favoured method of constraining the growth history is to measure the linear growth rate,
\be \label{eq:growthrate}
f(k, a) = \frac{d\log \Delta_M}{d \log a},
\ee
as a function of redshift and scale. Inserting (\ref{eq:growthrate}) into (\ref{eq:growth}), one obtains \citep{Baker:2013hia}
\bea \label{eq:growthode}
\frac{d f(k, a)}{d \log a} &=&  - f(k, a) \left [ f(k, a) + \frac{1}{2} - \frac{3}{2} w(a) [1 - \Omega_M(a)] \right ]  \nonumber\\
 & & ~ + \frac{3}{2} \Omega_M(a) \mu(k, a).
\eea
Equation (\ref{eq:growthode}) is readily solved using standard numerical techniques once $\mu(k, a)$ and $w(a)$ are given. While one might be tempted to try to preserve generality by keeping $\mu(k, a)$ completely arbitrary, large regions of this space of functions are likely to map onto contrived or physically implausible modified gravity theories. Instead, it is prudent to impose some structure, such as that suggested by the rather general Horndeski class of second-order scalar single-field modifications to GR \citep{1974IJTP...10..363H, 2011PhRvD..84f4039D}. These can be shown to restrict the form of $\gamma$ and $\mu$ to \citep{2011PhLB..706..123D, 2013PhRvD..87b3501A, 2013PhRvD..87j4015S}
\bea
\gamma(k, a) &=& \frac{p_1(a) + p_2(a) k^2}{1 + p_3(a) k^2}\\
\mu(k, a) &=& \frac{1 + p_3(a) k^2}{p_4(a) + p_5(a) k^2},
\eea
where $\{p_n\}$ are polynomials in scale factor only. This offers a strong restriction on the possible form of any scale dependence of the growth rate, but still allows five essentially arbitrary functions of time. Many choices of the polynomials result in theories that are unphysical (e.g. due to instabilities, tachyonic modes and so on), so the space of allowed models should be further restricted by applying viability conditions \citep[e.g.][]{2014JCAP...05..043P, 2015arXiv150603047P}. This is not simple in practise, so to make progress we follow a similar approach to \cite{2013MNRAS.429.2249S, Baker:2014zva, 2015arXiv150201590P} and apply the parametrisation
\be \label{eq:xiparam}
\mu(k, a) \approx 1 + A_\mu \frac{\Omega_\mathrm{DE}(a)}{\Omega_\mathrm{DE, 0}} \left [ 1 + \left ( \frac{k}{k_\mu}\right )^{-2}\right ].
\ee
Only two free parameters have been introduced here: $A_\mu$ sets the amplitude of the growth modifications ($A_\mu \to 0$ for GR), and $k_\mu$ sets the scale at which the growth begins to deviate from scale-independence. The dependence on $\Omega_\mathrm{DE}(a)$ recognises the motivation of modifications to GR as a possible explanation of cosmic acceleration, but alternative redshift dependences of the modifications can equally be considered. While there has clearly been some loss of generality, Eq.~(\ref{eq:xiparam}) at least attempts to account for redshift- and scale-dependent modifications to growth in a theoretically-motivated way \citep[particularly the possibility that MG would introduce a new physical scale, denoted by $k_\mu$;][]{Baker:2014zva}, and so we will use it as the preferred parametrisation here. Note that other methods of parametrising the growth equations have also been considered \citep[e.g.][]{2007PhRvD..76j4043H, 2010PhRvD..81l3508D}.

A number of phenomenological parametrisations are also in use throughout the literature. The simplest is \citep{peebles1980large, 2007APh....28..481L}
\be
f(a) = \Omega^\gamma_M(a),
\ee
where the $\gamma$ parameter is known as the growth index, equal to $0.55$ in $\Lambda$CDM+GR. Various extensions to this parametrisation have been proposed to capture modifications to the growth history in non-standard scenarios, including allowing the growth index to depend on redshift or the dark energy equation of state \citep[e.g.][]{2014JCAP...05..042S}, or by also modifying the overall amplitude of the growth rate \citep{PhysRevD.77.083508},
\be
f(a) = \Omega^\gamma_M(a) [1 + \eta(a)].
\ee
For ease of comparison with previous results, we include forecasts for a few variations on these parametrisations as well as for Eq.~(\ref{eq:xiparam}). In particular, we will consider simple redshift-dependent modifications to the growth index and amplitude of the growth rate, $\gamma(a) \approx \gamma_0 + \gamma_1 (1 - a)$ and $\eta(a) \approx \eta_0 + \eta_1 (1 - a)$.

The growth rate can be measured in a number of ways, mostly by using probes of the peculiar velocity field (since $v\!\! \propto\!\! f$ on linear, subhorizon scales) \citep[e.g.][]{2008PhRvD..78f3503J, Kosowsky:2009nc, Hellwing:2014nma, Mueller:2014nsa, 2015arXiv150406885J}. We will concentrate on just one method -- observations of redshift-space distortions (RSDs), the anisotropy induced in the galaxy correlation function in redshift-space by coherent peculiar velocities \citep{Kaiser:1987qv, 2011RSPTA.369.5058P}. While not as robust as BAOs in terms of insensitivity to complicating non-linear processes and other systematic effects \citep{2010PhRvD..81d3512S, 2013PhRvL.111p1301M}, one can at least model RSDs on large scales using linear cosmological perturbation theory plus small corrections, and they have already been successfully used to test GR \citep{2008Natur.451..541G, 2013MNRAS.429.1514S, 2014MNRAS.443.1065B}.

For small angular separations, the anisotropy induced in the redshift-space matter power spectrum is well described by the Kaiser approximation \citep{Kaiser:1987qv},
\bea
P(\mathbf{k}, z) &=& F_\mathrm{RSD}(\mathbf{k}, z) P(k, z) \\
F_\mathrm{RSD}(\mathbf{k}, z) &=& \left ( b(z, k) + f(z, k) \mu^2 \right )^2 e^{-k^2 \mu^2\sigma_\mathrm{NL}^2}, \label{eq:rsd}
\eea
where $\mu = \cos \theta$ is the angle of the wavevector to the line of sight, $b$ is the bias of the tracer population with respect to the dark matter, and an exponential term has been added to account for the smearing-out of redshift information on small scales by incoherent, non-linear peculiar velocities. The growth rate can be separated from the (generally poorly-known) bias by comparing moments of the redshift-space matter distribution. There is a degeneracy with the normalisation of the power spectrum, $\sigma_8$, however, such that only the combinations $b\sigma_8$ and $f\sigma_8$ (or alternatively $b\sigma_8$ and $\beta = f/b$) can be measured directly. This degeneracy can be broken using one of the parametric growth models from above, combined with a constraint on the normalisation of the power spectrum from the CMB.

\section{Forecasting for SKA HI surveys} \label{sec:fisher}

In this section, we describe the HI galaxy redshift and intensity mapping surveys that will be performed by the SKA, and outline the formalism used to forecast constraints on the expansion and growth rates and their parametrisations.

Galaxy redshift surveys are a tried and tested technique in the optical and near-infrared, where they have already been used to measure the BAO and RSDs to high precision at $z \lesssim 1$. Their survey speed scales poorly with increasing volume, however, as they rely on making time-consuming high-SNR detections of individual objects. Intensity mapping relaxes the requirement for individual detections, thus promising dramatically improved survey speeds. It is so far a relatively untested method however, with the first detection of the cosmological fluctuations having been made only recently with this technique \citep{2013ApJ...763L..20M}. Nevertheless, a host of IM experiments are planned to demonstrate the feasibility of the method over the next few years \citep{Bull:2014rha}, paving the way for the surveys with Phase 1 of the SKA that we consider here. We will consider the two methods on an equal footing.

\subsection{Fisher forecasting formalism} \label{sec:formalism}

Fisher forecasting is a simple, computationally inexpensive way of predicting the constraints on a set of parameters that should be achieved by a given experimental configuration. While clearly approximate and idealised -- it assumes Gaussianity and neglects systematic biases -- Fisher forecasting is nevertheless a reliable way of understanding the relative performance of different experiments and getting a handle on correlations between parameters.

To proceed, one must first define fiducial models for the expected signal and noise for a set of observations, as a function of the parameters of interest. Models for galaxy redshift and intensity mapping surveys are constructed in subsequent sections, based on the formalism developed in \cite{Bull:2014rha}. The Fisher matrix for a set of parameters $\{\theta\}$ can be written as \citep{2007ApJ...665...14S}
\be
F_{i j} = \int \frac{d^3k}{(2\pi)^3} V_\mathrm{eff}(\mathbf{k}) \frac{\partial \log C^S}{\partial \theta_i} \frac{\partial \log C^S}{\partial \theta_j}.
\ee
The effective volume,
\be
V_\mathrm{eff}(\mathbf{k}) = f_\mathrm{sky} V_i \left [\frac{C^S(\mathbf{k}, z)}{C^S(\mathbf{k}, z) + C^N(\mathbf{k}, z)} \right ]^2,
\ee
is a weighting that accounts for the varying sensitivity of an experiment to different Fourier modes (e.g. due to instrumental beam effects or cosmic variance), and depends on the physical volume of the redshift bin, $V_i = \int_{z_\mathrm{min}}^{z_\mathrm{max}} (dV/dz) dz$, and the fraction of the sky covered by the survey, $f_\mathrm{sky} = S_\mathrm{area} / 4\pi$. {\corr The $C^S(\mathbf{k}, z)$ and $C^N(\mathbf{k}, z)$ terms are the signal and noise covariance respectively, and will be explicitly defined in subsequent sections.} After calculating the Fisher matrix, one can invert it to get an estimate of the expected covariance between the parameters, $F^{-1} \approx \mathrm{Cov}(\{\theta\})$.

\begin{figure}[t]
\hspace{-1em}\includegraphics[width=1.04\columnwidth]{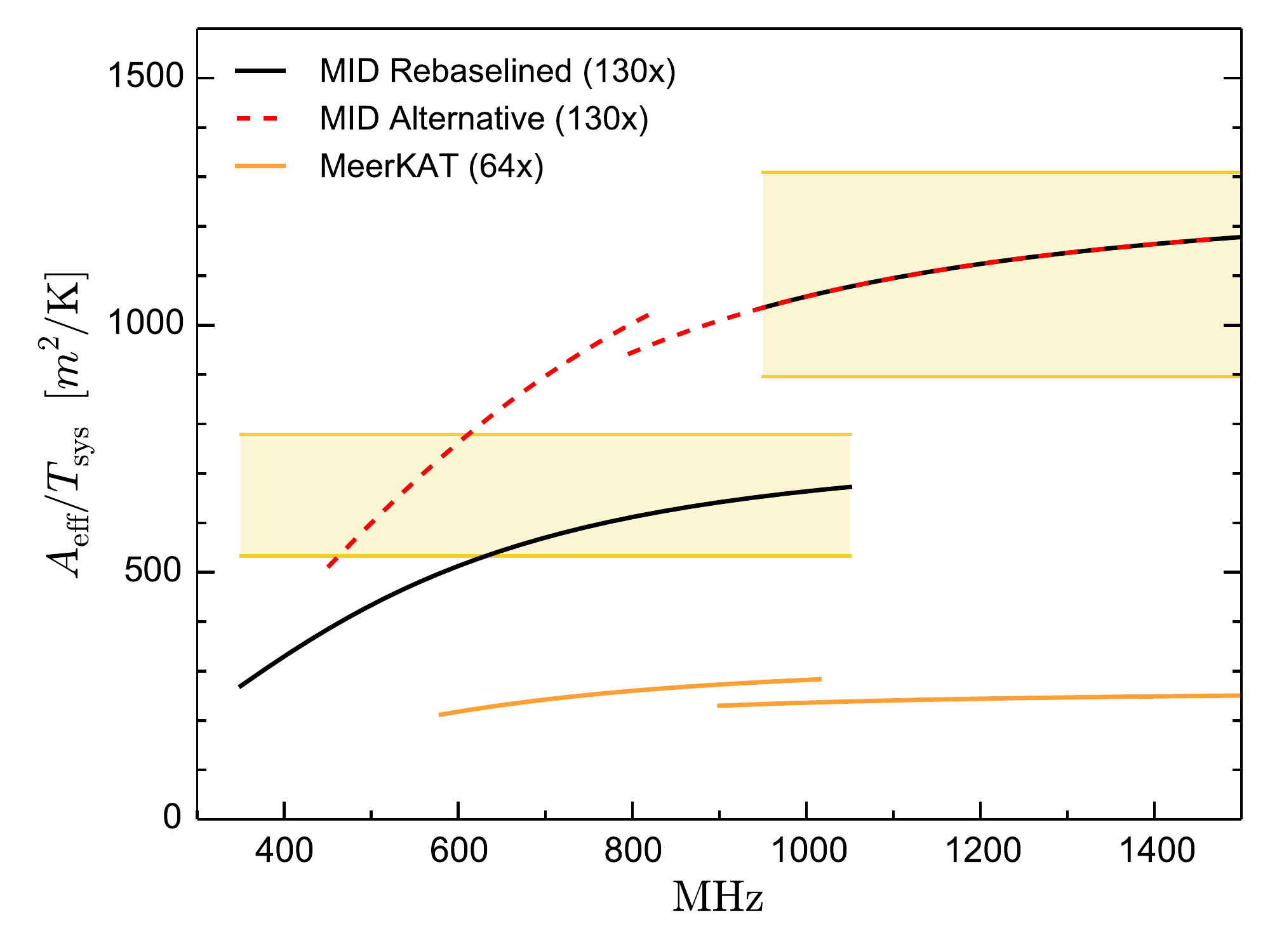}
\caption{The sensitivity of the `Rebaselined' and `Alternative' designs for the SKA1-MID receivers, shown as total $A_{\rm eff} / T_{\rm sys}$ curves for the sub-array. The MeerKAT bands (orange lines) are shown for comparison. The yellow shaded regions denote the SKA1 Baseline specification sensitivity \citep{SKAbaseline} for the original assumption of 190 dishes (upper limit), and corrected for the post-rebaselining figure of 130 dishes (lower limit).}
\label{fig:midsens}
\end{figure}

\begin{table*}[t]
\begin{center}
{\renewcommand{\arraystretch}{1.9} \begin{tabular}{|ll|r|r|r|r|r|r|r|r|c|c|c|c|c|}
\hline 
\multirow{2}{*}{\bf Telescope array} & & $T_\mathrm{recv}$ & $A_\mathrm{eff}$~ & $D_\mathrm{dish}$ & \multirow{2}{*}{$N_\mathrm{d} \times N_\mathrm{b}$~} & ${\rm FOV}$~ & $\nu_{\rm ref}$~~ & $\nu_\mathrm{min}$~ & $\nu_\mathrm{max}$~ & \multirow{2}{*}{$z_\mathrm{min}$} & \multirow{2}{*}{$z_\mathrm{max}$} & $S_{\rm area}$ IM & $S_{\rm area}$ GS \\
 & & [K]~~ & [m$^2$] & [m]~~ & & [deg$^2$] & [MHz] & [MHz] & [MHz] & &  & [deg$^2$] & [deg$^2$] \\
\hline
\multirow{2}{*}{SKA1-LOW}
 & Rebaselined & 40~~~ & 925~ & 35~~~~ & $455\times3$ & 28~~~~~ & 110~~ & 50~~ & 350~ & 3.06 & 27.4 & \multirow{2}{*}{1,000} & \multirow{2}{*}{---} \\
 & Alternative & 40~~~ & 925~ & 35~~~~ & $455\times3$ & 28~~~~~ & 110~~ & 50~~ & 500~ & 1.84 & 27.4 & & \\
\hline
\multirow{2}{*}{SKA1-MID B1}
 & Rebaselined & 23~~~ & 133~ & 15~~~~ & $130\times1$ & 1.78 & 700~~ & 350~~ & 1050~ & 0.35 & 3.06 & \multirow{3}{*}{25,000} & \multirow{3}{*}{---} \\
 & Alternative & 12~~~ & 133~ & 15~~~~ & $130\times1$ & 1.78 & 700~~ & 450~~ & 825~ & 0.72 & 2.16 & & \\
MeerKAT & UHF-band & 23~~~ & 115~ & 13.5~ & $64\times1$ & 2.20 & 700~~ & 580~~ & 1015~ & 0.40 & 1.45 & & \\
\hline
\multirow{2}{*}{SKA1-MID B2}
 & Rebaselined & 15.5 & 150~ & 15~~~~ & $130\times1$ & 0.87 & 1000~~ & 950~~ & 1760~ & 0.00 & 0.49 & \multirow{3}{*}{25,000} & \multirow{3}{*}{5,000} \\
 & Alternative & 15.5 & 150~ & 15~~~~ & $130\times1$ & 0.87 & 1000~~ & 795~~ & 1470~ & 0.00 & 0.79 & & \\
MeerKAT & L-band & 30~~~ & 122~ & 13.5~ & $64\times1$ & 1.08 & 1000~~ & 900~~ & 1670~ & 0.00 & 0.58 & & \\
\hline
SKA2 & & 15~~~ & 6~ & 3.1~ & $70000\times1$ & 30~~~~~ & 1000~~ & 470~~ & 1290~ & 0.10 & 2.00 & --- & 30,000 \\
\hline
\end{tabular} }
\end{center}
\caption{Representative instrumental parameters for various assumed SKA Phase 1 and 2 array configurations and bands. $A_{\rm eff}$ is the effective area per dish or receiving element, and the ${\rm FOV}$ is calculated at {\corr the reference frequency, $\nu_{\rm ref}$. $\nu_{\rm min, max}$ are the band edges, and $z_{\rm min, max}$ the corresponding minimum/maximum accessible redshifts. Representative survey areas for intensity mapping (IM) and galaxy surveys (GS) are given in the final two columns. Other quantities are defined in the text.}}
\label{tab:skaarrays}
\end{table*}

\subsection{SKA array configurations}
\label{sec:skaconfig}

The design of the SKA has not yet been finalised, so there is some freedom in what to assume for the instrumental specifications. The most complete specification for Phase 1 of the SKA is currently the `baseline' design of \cite{SKAbaseline}. This originally called for a three-array system, consisting of two dish arrays with mid-frequency receivers covering multiple bands (SKA1-MID and SUR) and a single low-frequency aperture array (LOW). This specification has now been updated following a `rebaselining' procedure however \citep{SKArebaselining}, which has removed the SUR array, halved the number of receiving stations of LOW, and reduced the number of MID dishes by 30\% in order to meet a cost cap.

Following this, options for redefining the available MID and LOW bands are also being considered, with the intention of better-aligning the frequency coverage with recently-selected `key science' goals \citep{SKAlevel0}. This is also an opportunity to improve the overall sensitivity of the array: by reducing the bandwidth of Band 1 of MID, simpler low-noise receivers (with a maximum:minimum frequency ratio of 1.85:1 or less) can be used instead of the more complex, higher-noise wideband (3:1 ratio) receivers of the current design, which also suffer from degraded performance at lower frequencies. Fig.~\ref{fig:midsens} shows a measure of sensitivity -- the total effective area ($A_{\rm eff}$) of the array divided by the system temperature ($T_{\rm sys}$) -- for the original baseline specification, the most up-to-date sensitivity estimates following rebaselining \citep{SKAsens}, and a proposed `alternative' Band 1 and 2 configuration with simpler, reduced-bandwidth receivers (J. Jonas \& M. Santos, priv. comm.). The alternative configuration almost doubles $A_{\rm eff}/T_{\rm sys}$ at 800 MHz, mitigating the cuts from rebaselining. The maximum frequency of LOW could be increased to 500 MHz to compensate for the resulting loss of frequency coverage of Band 1, without affecting the sensitivity of LOW in its design range of $50 \le \nu \le 350$ MHz (K. Zarb-Adami, priv. comm.).

To reflect the fluidity of the design, we therefore chose two representative configurations for the SKA1-LOW and MID arrays: a `Rebaselined' design based on the recommendations of \cite{SKArebaselining} and using the updated sensitivity estimates of \cite{SKAsens}; and an `Alternative', also based on the rebaselined specification, but with a different set of frequency bands and reduced system temperature to reflect the expected improvement in sensitivity if simpler receivers can be used. In our forecasts for SKA1-MID we also include the 64 MeerKAT\footnote{\url{http://public.ska.ac.za/meerkat}} dishes, which significantly increases the sensitivity over part of each band (see Fig.~\ref{fig:midsens}). The specifications are summarised in Table \ref{tab:skaarrays}.

The design for Phase 2 of the SKA is yet to be formally defined, and only notional specifications currently exist.\footnote{\url{http://astronomers.skatelescope.org/ska2/}} The expectation is that it will have approximately $10\times$ the sensitivity of SKA1 above 350 MHz, and will be capable of (spectroscopically) detecting $\sim10^9$ HI galaxies. In this paper, we assume a dense aperture array design \citep[c.f.][]{2015arXiv150403854T} with the frequency range used in \cite{2015MNRAS.450.2251Y}, and choose the collecting area and number of receivers such that it would detect $\sim10^9$ HI galaxies above a $10\sigma$ threshold (see Sect. \ref{sec:fishergal}). The resulting specification is listed in Table \ref{tab:skaarrays}. Note that this actually leads to a total $A_{\rm eff} / T_{\rm sys}$ $\approx 15\times$ that of SKA1-MID + MeerKAT at 1 GHz.

\subsection{Galaxy redshift surveys} \label{sec:fishergal}

Galaxy surveys detect and then measure the positions and redshifts of many individual galaxies, which are assumed to Poisson sample the underlying dark matter density field. The measured galaxy correlation function can then be used to infer the matter power spectrum in redshift-space, which contains a host of information about the growth of structure and other cosmological quantities, as discussed above.

The signal covariance for a galaxy redshift survey is simply the redshift-space power spectrum,
\be \label{eq:CS_galaxy}
C^S = P(\mathbf{k}, z) = F_\mathrm{RSD}(\mathbf{k}, z) P(k),
\ee
where $F_\mathrm{RSD}$ was defined in Eq.~(\ref{eq:rsd}), and we marginalise over the non-linear velocity dispersion scale, $\sigma_\mathrm{NL} \approx 7$ Mpc, as a nuisance parameter. The noise covariance is modelled as shot noise, $C^N = 1/n(z)$. In the limit that $n(z) \to \infty$, $V_\mathrm{eff} \to f_\mathrm{sky} V_i$, and the Fisher errors are limited only by the finite number of Fourier modes in the survey volume (i.e. sample variance). We set $k_{\rm max} = 0.14$ Mpc$^{-1} (1+z)^{2/(2+n_s)}$, corresponding to a non-linear cutoff on small scales \citep{2003MNRAS.341.1311S}.

The number density and bias functions depend on the type of galaxy being targeted by the survey, as well as the sensitivity of the survey instrument. The rms flux sensitivity for dual-polarisation receivers is given by
\be
S_{\rm rms} = \frac{2 k_B T_{\rm sys}}{A_{\rm eff} N_{\rm dish} \sqrt{2 \delta\nu\, t_p}},
\ee
where $t_p$ is the integration time per pointing, $N_{\rm dish}$ and $A_{\rm eff}$ are the number and effective area of the collecting elements, $\delta \nu$ is the channel bandwidth, and the total system temperature is $T_\mathrm{sys} = T_\mathrm{recv} + T_\mathrm{sky}$, where the first term is the total contribution of the receiver system to the noise, and the second is due to background emission, $T_\mathrm{sky} \approx 60\, \mathrm{K} \times (\nu / 300 \mathrm{MHz})^{-2.5}$. For all array configurations, we assume a per-element collecting area of $A_{\rm eff} \approx \epsilon \pi (D_{\rm dish}/2)^2$, where $D_{\rm dish}$ is the dish/station diameter and $\epsilon \approx 0.7-0.9$ is a typical aperture efficiency. The integration time per pointing can be rewritten in terms of the total survey time, $t_p = t_{\rm tot}\, {\rm FOV} / S_{\rm area}$, where the instantaneous field of view is ${\rm FOV} \approx \frac{\pi}{8} (1.3\lambda / D_{\rm dish})^2$.

The number density and bias of HI galaxies as a function of redshift and flux sensitivity were calculated by \cite{2015MNRAS.450.2251Y} using the S-cubed simulations \citep{2009ApJ...703.1890O}, which are based on the Millennium dark matter only simulation \citep{2005Natur.435..629S}. A frequency resolution of 10 kHz and detection threshold of $10\sigma$ was assumed in these calculations, and the bias was calculated using the halo model with a mass function measured from the simulations. We calculate the number density and bias for the array configurations in Table \ref{tab:skaarrays} by interpolating the results of \cite{2015MNRAS.450.2251Y} for 10,000 hour surveys over either 5,000 deg$^2$ (SKA1) or 30,000 deg$^2$ (SKA2). The dependence of survey performance on the assumed survey time and area is discussed in Appendix \ref{app:survey}, and fitting functions for the number density and bias are given in Appendix \ref{app:dndz}. To boost the number density of detected sources, we also revise the detection threshold for SKA1 to $5\sigma$, which is relatively low \citep[c.f.][]{2005MNRAS.360...27A}. Only Band 2 of MID is usable; the number densities obtained with Band 1 and LOW would be too small for a sufficiently large survey area and reasonable survey time. We have also assumed a source detection efficiency of 100\%, but note that effects like source confusion will reduce this \citep{2015MNRAS.449.1856J}. The resulting values of $n(z)$ and $b(z)$ in bins of width $\Delta z = 0.1$ are listed in Table \ref{tab:nzska1} (SKA1) and Table \ref{tab:nzska2} (SKA2).

\begin{table}[t]
\begin{center}
{\renewcommand{\arraystretch}{1.6}
\begin{tabular}{c|c|c|ccr|}
\cline{2-6}
 & \multicolumn{2}{c|}{} & \multicolumn{3}{c|}{\bf SKA1-MID B2 + MeerKAT} \\
\cline{2-6}
 & $z_{\rm min}$ & $z_{\rm max}$ & $n(z)$ [Mpc$^{-3}$] & $b(z)$ & $S_{\rm rms}$ $[\mu{\rm Jy}]$ \\
\cline{1-6}
\parbox[t]{2mm}{\multirow{5}{*}{\rotatebox[origin=c]{90}{Rebase. / Alt.}}} & 0.0 & 0.1 & $2.73 \times 10^{-2}$ & 0.657 & 117.9~~~~ \\
 & 0.1 & 0.2 & $4.93 \times 10^{-3}$ & 0.714 & 109.6~~~~ \\
 & 0.2 & 0.3 & $9.49 \times 10^{-4}$ & 0.789 & 102.9~~~~ \\
 & 0.3 & 0.4 & $2.23 \times 10^{-4}$ & 0.876 & 97.5~~~~ \\
 & 0.4 & 0.5 & $6.44 \times 10^{-5}$ & 0.966 & 93.1~~~~ \\
\cline{1-6}
\parbox[t]{2mm}{\multirow{2}{*}{\rotatebox[origin=c]{90}{Alt.}}} & 0.5 & 0.6 & $1.70 \times 10^{-5}$ & 1.059 & 89.4~~~~ \\
 & 0.6 & 0.7 & $1.72 \times 10^{-6}$ & 1.208 & 106.2~~~~ \\
\cline{1-6}
\end{tabular}}
\caption{Binned number density and bias of HI galaxies, and corresponding flux sensitivity, for SKA1-MID Band 2, assuming a 5,000 deg$^2$, 10,000 hour survey. The detection threshold is 5$\sigma$.} \vspace{-2em}
\label{tab:nzska1}
\end{center}
\end{table}

\subsection{Intensity mapping surveys} \label{sec:fisherim}

The process of cataloguing millions of galaxies over a large volume is extremely time consuming, as each source must be detected at a sufficiently high signal-to-noise ratio (SNR) to ensure that it is not just a statistical fluctuation. An alternative approach is to make low-resolution maps of the integrated emission from many galaxies, in a process {\corr known as {\it intensity mapping} \citep{1997ApJ...475..429M, 2001JApA...22...21B, 2004MNRAS.355.1339B, 2008PhRvL.100p1301L, 2008PhRvL.100i1303C}}. Fluctuations in the observed intensity of redshifted HI emission follow fluctuations in the underlying matter density field as traced by the HI emitting galaxies, allowing the density field to be reconstructed on sufficiently large scales from intensity maps. Redshift information is included automatically as the target is an emission line {\corr (i.e. the frequency of observation maps directly to the redshift of emission, $\nu_{\rm obs} = \nu_{\rm 21cm} / (1+z)$)}, and one only need integrate down to $\mathrm{SNR} \approx 1$ to {\corr (statistically)} detect the cosmological signal (c.f. measurement of CMB anisotropies). Intensity mapping surveys are therefore more rapid, making it possible to probe larger volumes in a reasonable amount of time.

\begin{table}[t]
\begin{center}
{\renewcommand{\arraystretch}{1.6}
\begin{tabular}{|c|c|ccc|}
\hline
\multicolumn{2}{|c|}{} & \multicolumn{3}{c|}{\bf SKA2} \\
\hline
$z_{\rm min}$ & $z_{\rm max}$ & $n(z)$ [Mpc$^{-3}$] & $b(z)$ & $S_{\rm rms}$ $[\mu{\rm Jy}]$ \\
\hline
0.1 & 0.2 & $6.20\times 10^{-2}$ & 0.623 & 6.1 \\
0.2 & 0.3 & $3.63\times 10^{-2}$ & 0.674 & 5.7 \\
0.3 & 0.4 & $2.16\times 10^{-2}$ & 0.730 & 5.4 \\
0.4 & 0.5 & $1.31\times 10^{-2}$ & 0.790 & 5.2 \\
0.5 & 0.6 & $8.07\times 10^{-3}$ & 0.854 & 5.0 \\
0.6 & 0.7 & $5.11\times 10^{-3}$ & 0.922 & 4.8 \\
0.7 & 0.8 & $3.27\times 10^{-3}$ & 0.996 & 4.7 \\
0.8 & 0.9 & $2.11\times 10^{-3}$ & 1.076 & 4.6 \\
0.9 & 1.0 & $1.36\times 10^{-3}$ & 1.163 & 4.6 \\
1.0 & 1.1 & $8.70\times 10^{-4}$ & 1.257 & 4.5 \\
1.1 & 1.2 & $5.56\times 10^{-4}$ & 1.360 & 4.5 \\
1.2 & 1.3 & $3.53\times 10^{-4}$ & 1.472 & 4.5 \\
1.3 & 1.4 & $2.22\times 10^{-4}$ & 1.594 & 4.5 \\
1.4 & 1.5 & $1.39\times 10^{-4}$ & 1.726 & 4.5 \\
1.5 & 1.6 & $8.55\times 10^{-5}$ & 1.870 & 4.5 \\
1.6 & 1.7 & $5.20\times 10^{-5}$ & 2.027 & 4.5 \\
1.7 & 1.8 & $3.12\times 10^{-5}$ & 2.198 & 4.6 \\
1.8 & 1.9 & $1.83\times 10^{-5}$ & 2.385 & 4.6 \\
1.9 & 2.0 & $1.05\times 10^{-5}$ & 2.588 & 4.7 \\
\hline
\end{tabular}}
\caption{Number density and bias of HI galaxies for SKA2, assuming a 30,000 deg$^2$ survey for 10,000 hours, and a 10$\sigma$ threshold.} 
\label{tab:nzska2}
\end{center}\vspace{-2em}
\end{table}

The signal covariance for the HI brightness temperature fluctuations measured by an intensity mapping survey is
\be
C^S(\mathbf{k}, z) = T_b^2 F_\mathrm{RSD}(\mathbf{k}, z) P(k),
\ee
where the only difference with Eq.~(\ref{eq:CS_galaxy}) is a factor of the HI brightness temperature squared, $T^2_b(z)$. We use the bias and brightness temperature models from \cite{Bull:2014rha}. The noise covariance for an IM survey can be written as a scale-dependent effective `number density' \citep{Bull:2014rha},
\bea \label{eq:nIM}
n_\mathrm{IM}(\mathbf{k}, z) &=& N_\mathrm{dish}\; N_\mathrm{beam}\; N_\mathrm{pol} \nonumber\\
 &\times& \left ( \frac{T_b}{T_\mathrm{sys}}\right )^2 \frac{\epsilon^2 \,t_\mathrm{tot}}{S_\mathrm{area} \nu_\mathrm{HI}} B_\perp^2 B_\parallel.
\eea
This expression is based on the ideal radiometer equation. The factors of $B_\perp$ and $B_\parallel$ are window functions in Fourier space due to the angular beam and finite frequency channel bandwidth, and $N_\mathrm{dish}$, $N_\mathrm{beam}$, and $N_\mathrm{pol}$ are factors that account for the gain in survey speed for a telescope array with multiple dishes/beams/polarisation channels. The noise also depends on the survey area, $S_\mathrm{area}$, and the total observation time for the survey, $t_\mathrm{tot}$. See Table \ref{tab:skaarrays} for the values of all relevant parameters for the various SKA configurations.

Eq.~(\ref{eq:nIM}) depends on the type of receiver system through the $B_\perp$ term. We have assumed that the SKA1-MID IM surveys will be performed in autocorrelation mode; in interferometer mode, MID has too small a field of view and an insufficient density of short baselines to be sensitive to the large scales we are interested in at redshifts of $z \lesssim 1.4$ \citep{Bull:2014rha}. Its interferometric resolution is better matched to (e.g.) the BAO scale at higher redshifts, however. Autocorrelation observations are sensitive to all scales between approximately the size of the beam and the survey area, so are better matched to the BAO scale at lower redshift for MID. They suffer from correlated ($1/f$) noise and ground pickup however, which can significantly increase the difficulty of recovering the cosmological signal \citep[although foreground cleaning and appropriate scanning strategies can help to remove these effects; see][]{2015arXiv150704561B}. SKA1-LOW is an aperture array, which operates as an interferometer and has a configuration that is better-matched to the BAO scale at high $z$. Noise and beam expressions for the various types of receiver, including baseline density distributions for the interferometers, are given in \cite{Santos:2015bsa} and \cite{Bull:2014rha}.

Another important systematic effect is the presence of foreground contamination. Galactic synchrotron and other foregrounds are around 5--6 orders of magnitude brighter than the cosmological HI signal, and so must be removed with a high level of efficiency. Most foregrounds should be spectrally smooth, making it possible to subtract them using polynomial fitting, Principal Component Analysis, or similar \citep{2006ApJ...650..529W, 2006ApJ...648..767M, 2011PhRvD..83j3006L, 2011MNRAS.413.2103P, Alonso:2014ywa}. The frequency dependence of the beam response can hinder this process, however, with interferometers in particular susceptible to the generation of non-smooth foreground signals due to chromatic/wide-field effects \citep[the `foreground wedge':][]{2009ApJ...695..183B, 2009MNRAS.394.1575L, 2010ApJ...724..526D, 2012ApJ...752..137M, 2015ApJ...804...14T, 2015arXiv150806503S}. Other effects, such as atmospheric noise \citep{2015arXiv150704561B}, ionospheric distortions, and radio-frequency interference \citep{Alonso:2014sna}, can also be problematic at low and high frequencies respectively. Nevertheless, recent simulation work has shown that existing foreground removal methods can recover the true HI power spectrum to within $5\%$, although over-subtraction of the HI signal biases the recovered spectrum in a scale-dependent way \citep{2014MNRAS.441.3271W, Alonso:2014sna, 2015arXiv150704561B, 2015arXiv150900742O}. This is problematic if the aim is to use the broadband shape of the power spectrum for cosmology; recovery of the BAO scale is not biased by this effect, however \citep{2014MNRAS.441.3271W, Alonso:2014sna}.

We assume $t_{\rm tot} = 10^4$ hours for all IM surveys. The survey area is taken to be 25,000 deg$^2$ for MID, and 1,000 deg$^2$ for LOW. The MID results do not depend strongly on the assumed survey area for $S_{\rm area} \gtrsim 5,000$ deg$^2$; see Appendix \ref{app:survey}.

\subsection{Prior information} \label{sec:priorinfo}

It is useful to include prior information from other sources in the forecasts, e.g. in order to break degeneracies. By the time of the first SKA HI surveys in the early 2020's, a large amount of precision data from various sources will already be available. Rather than trying to forecast for the entire state of observational cosmology at that time, we take a more conservative approach and restrict the prior information to just two sources: the CMB angular power spectrum from Planck, and galaxy clustering information from BOSS, which anchor the constraints at high- and low-redshift respectively. For Planck, we use the DETF Fisher matrix prior, calculated assuming full polarisation, 80\% sky coverage, and 3 frequency bands free of foreground contamination \citep{Albrecht:2009ct}. For BOSS, we take the binned number density and bias values from \cite{Font-Ribera:2013rwa} for a 10,000 deg$^2$ survey and perform our own forecasts using the procedure outlined in Section \ref{sec:fishergal}. The priors are applied by adding the Planck and BOSS Fisher matrices to the Fisher matrix for a given SKA survey.

\subsection{Parameters used in the forecasts} \label{sec:params}

In all cases we forecast for the parameters
\be
\{ D_{\rm A}(z), H(z), f\sigma_8(z), b\sigma_8(z), \sigma_\mathrm{NL} \}, \nonumber
\ee
where $D_{\rm A}$ is the angular diameter distance and $H$ is the expansion rate. The first 4 parameters are assumed to be free in each redshift bin, and the non-linear velocity dispersion, $\sigma_\mathrm{NL}$, is marginalised as a nuisance parameter. This set of parameters can be viewed as ``model-independent'', as we have not assumed parametric functional forms for any of the first four functions. No priors are applied to this set.

\begin{figure}[b]
\hspace{1em}\includegraphics[width=0.95\columnwidth]{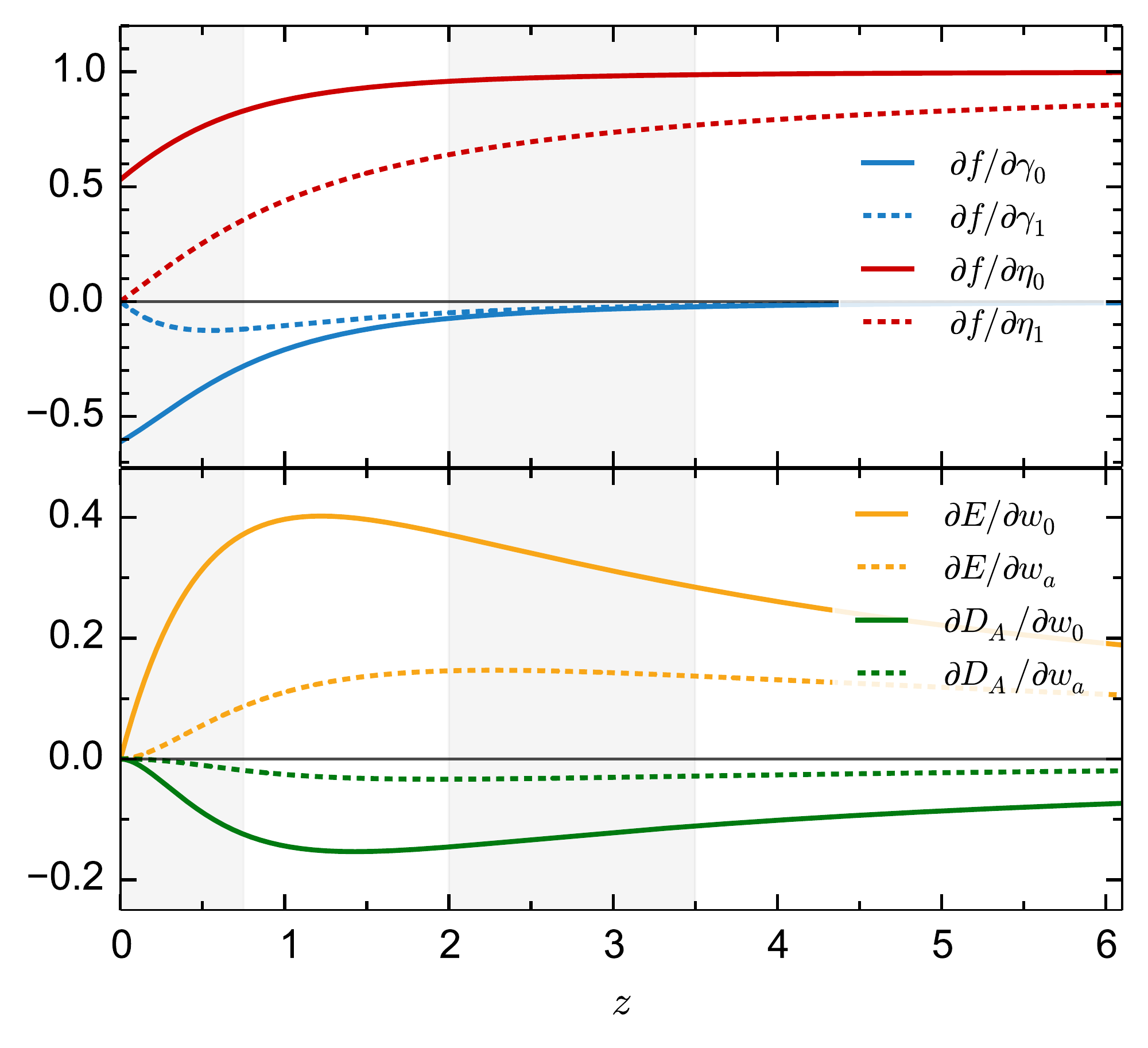}
\caption{Derivatives of $f(z)$, $E\! =\! H(z)/H_0$, and $D_{\rm A}(z)$ with respect to the modified growth and equation of state parameters. The $D_{\rm A}$ curves have been rescaled by a factor of $2 H_0 / c$.}
\label{fig:derivs}
\end{figure}

For the forecasts where a growth parametrisation is assumed, we project from the base parameters to
\be
\{ h, \Omega_\mathrm{DE}, \Omega_\mathrm{K}, \Omega_b h^2, w_0, w_a, n_s, \sigma_8, b(z), \sigma_\mathrm{NL} \} + \{\theta_\mathrm{MG}\}, \nonumber
\ee
where $\{\theta_\mathrm{MG}\}$ are growth parameters from one of the parametrisations discussed in Sect. \ref{sec:growth}. Both the BOSS and Planck priors are applied to this set, and the bias is marginalised over as a free parameter in each redshift bin. With a parametric model chosen for $f(z)$, {\corr the degeneracy between the bias and normalisation of the power spectrum that occurs in the RSD term is now broken, as the functional form of $\sigma_{8}(z)$ can be calculated from the growth model, and its normalisation is set by the measured CMB normalisation.}\footnote{For IM surveys, these quantities are also degenerate with $T_b(z)$. We assume that this is already known, and can be fixed in our analysis.} {\corr Explicitly,} we write $\sigma_8(z) = \sigma_8 D(z)$, where $\sigma_8\!\equiv\!\sigma_8(z\!\!=\!\!0)$ is now a separate parameter and $D(z)$ is the linear growth factor. Since $D(z)$ depends on $f$ through the definition $f = d\log D/d\log a$, its derivatives with respect to the growth parameters must also be taken into account. We do this by projecting $f\sigma_8$ into $\sigma_8$ and the growth (and CPL) parameters. Derivatives of $b\sigma_8$ with respect to these parameters are neglected in the projection, as this quantity would not be used to constrain $\{\theta_{\rm MG}\}$ in a realistic analysis. We continue to marginalise over the bias by projecting $b\sigma_8 \to b$ only.

\begin{figure*}[t]
\centering{
\includegraphics[width=1.04\columnwidth]{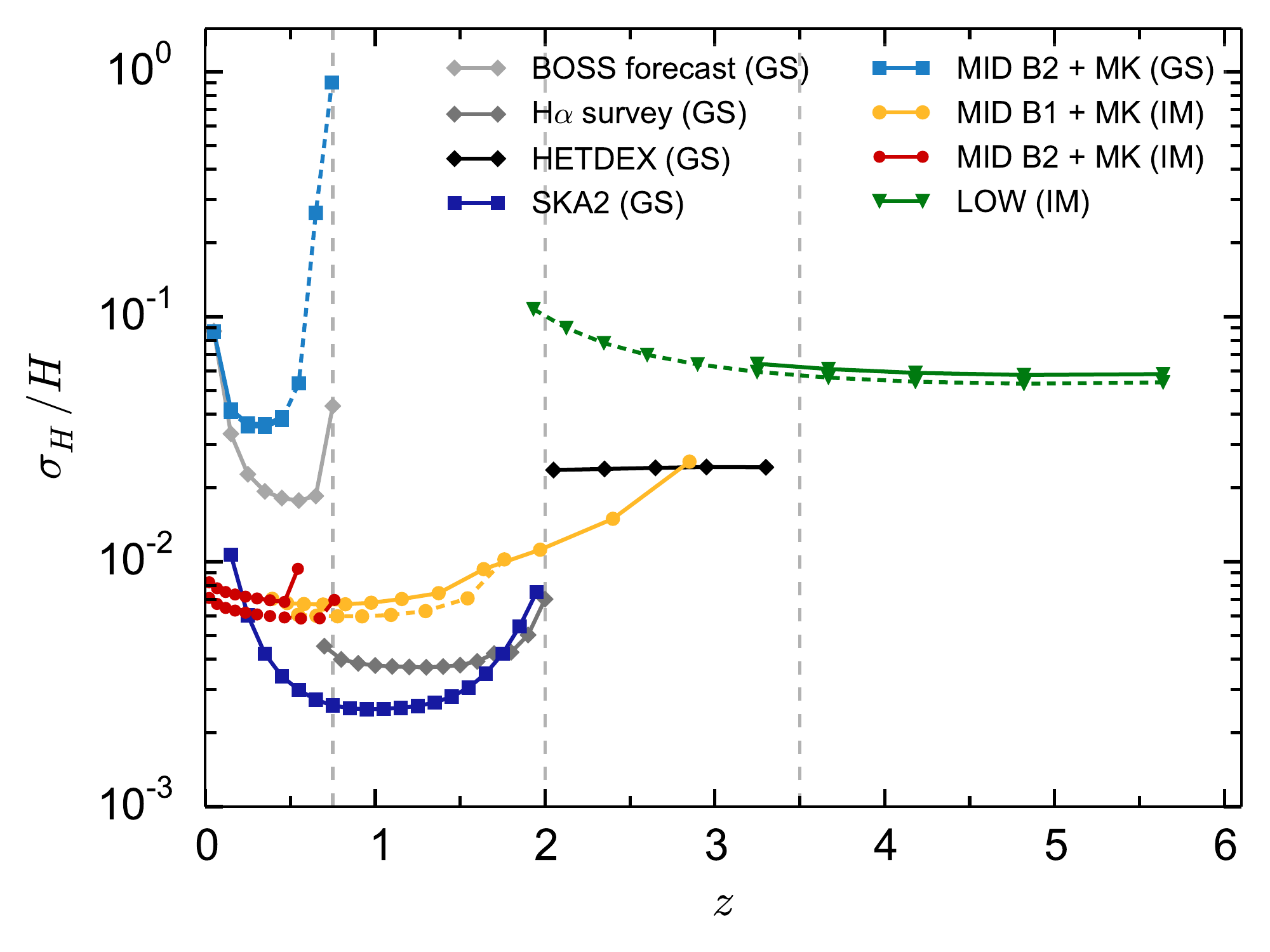}
\includegraphics[width=1.04\columnwidth]{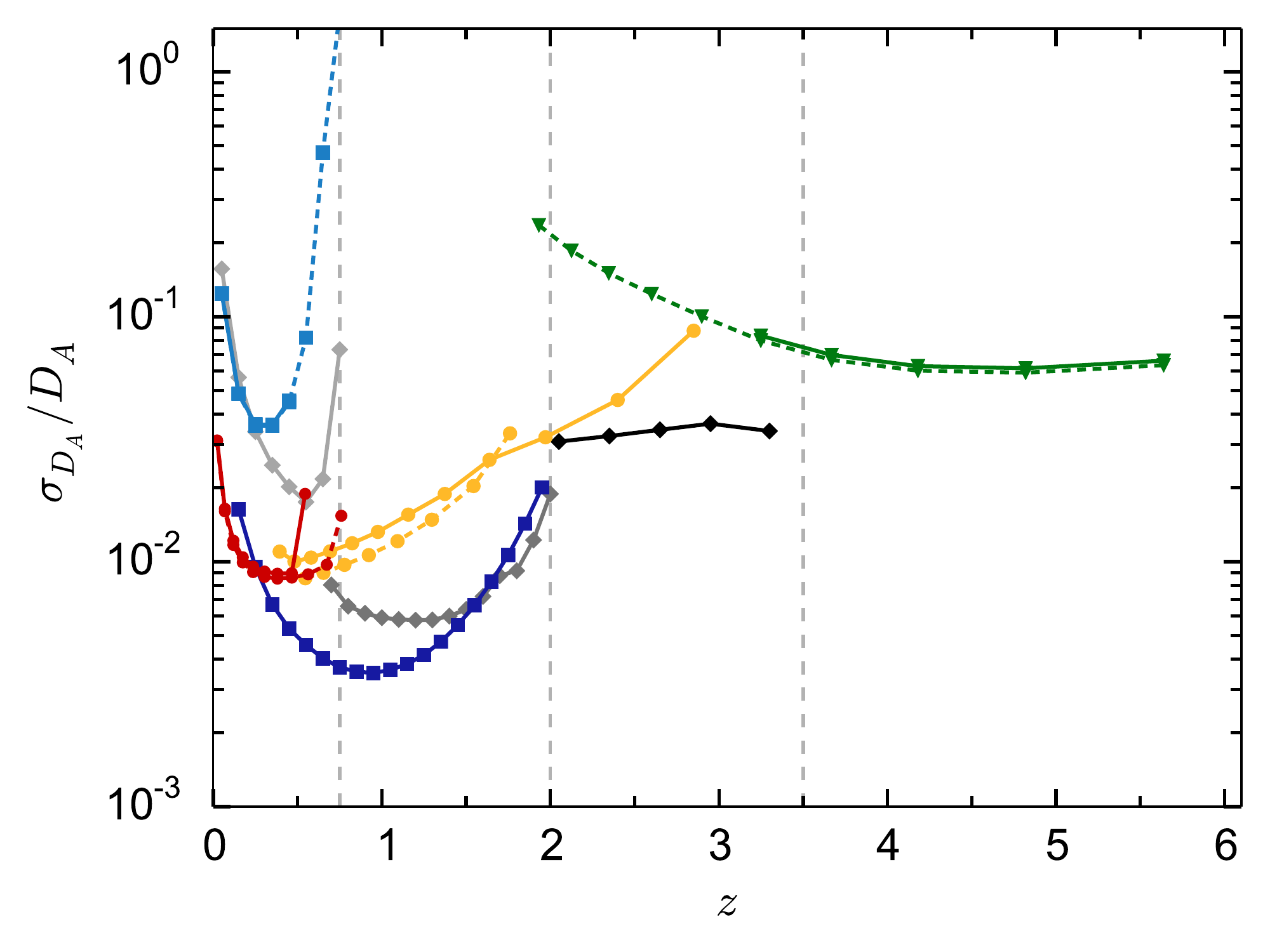}
\caption{Forecast constraints on $H(z)$ and $D_{\rm A}(z)$ for various galaxy surveys (GS) and intensity mapping surveys (IM). Dashed lines show the Alternative array configurations. The H$\alpha$ survey, BOSS and HETDEX specifications were taken from \cite{2013LRR....16....6A} and \cite{Font-Ribera:2013rwa} respectively.}
\label{fig:Hz}}
\end{figure*}


In all forecasts, we assume that information about the full shape of the power spectrum can be reliably recovered. Whether this is possible will depend on having sufficient control over scale-dependent systematic effects (e.g. see Sect. \ref{sec:fisherim}). A more conservative approach would be to use only RSDs and BAOs, as the latter are very robust to systematics \citep{Seo:2009fp, Mehta:2011xf}.

Finally, we use the \cite{2015arXiv150201589P} flat $\Lambda$CDM cosmology as our fiducial model, $(\Omega_M, \Omega_b, h, n_s, \sigma_8) = (0.316, 0.049, 0.67, 0.962, 0.834)$, unless otherwise specified.

\section{Results} \label{sec:results}

In this section we present forecast constraints on the background expansion rate, linear growth rate, and associated expansion and growth parametrisations.

To frame the discussion, it is first useful to divide the post-reionisation Universe into four approximate redshift regimes, each of which has something different to say about possible deviations from GR+$\Lambda$CDM. To help illustrate the differences between the regimes, Fig.~\ref{fig:derivs} shows the derivatives of $f(z)$, $H(z)$, and $D_{\rm A}(z)$ with respect to the expansion/growth parameters as a function of redshift for the fiducial cosmology. For a given measurement precision, measurements of the growth and expansion rates should be more sensitive to these parameters where the derivatives are largest, although degeneracies with other parameters strongly influence the constraints as well. The four regimes are as follows:
\paragraph{Low redshifts $(0 \le z \lesssim 0.7)$} The expansion becomes dark energy-dominated and the linear growth rate deviates appreciably from unity here, making this the most obvious regime in which to look for modifications to GR that are motivated as an explanation for cosmic acceleration. Fig.~\ref{fig:derivs} suggests that this redshift range should be the most sensitive to the growth index parameters, $\gamma_0$ and $\gamma_1$. Existing surveys have already made $\sim\,$few percent-level measurements of $D_{\rm A}(z)$ and $H(z)$ ($\sim\! 10\%$ for $f\sigma_8$) at these redshifts which, in combination with the CMB, constitute some of the most stringent cosmological constraints on GR to date \citep{2015arXiv150201590P}.

\paragraph{Dark energy transition $(0.7 \lesssim z \lesssim 2)$} This is where dark energy does not yet dominate the expansion but has started to become dynamically important, making it a suitable regime to look for evolution of the equation of state and growth index. This is backed up by Fig.~\ref{fig:derivs}, which shows that the derivatives peak in this range for many of the parameters. Precision measurements of expansion and growth have only reached the lower part of this range so far, but it is an important focus for a number of forthcoming surveys like Euclid and DESI.

\begin{figure*}[t]
\vspace{1em}\centering{
\includegraphics[width=1.02\columnwidth]{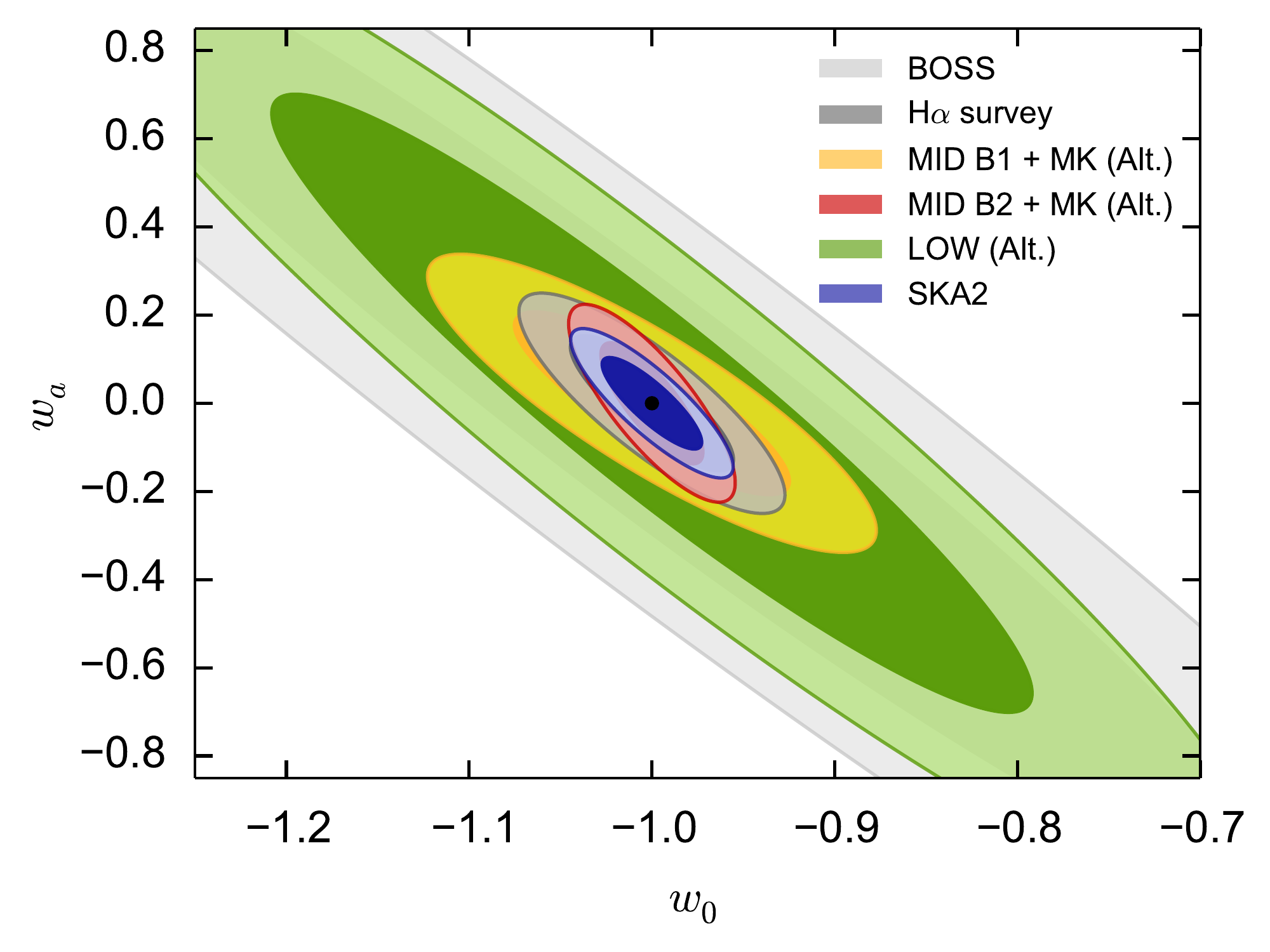}
\includegraphics[width=1.04\columnwidth]{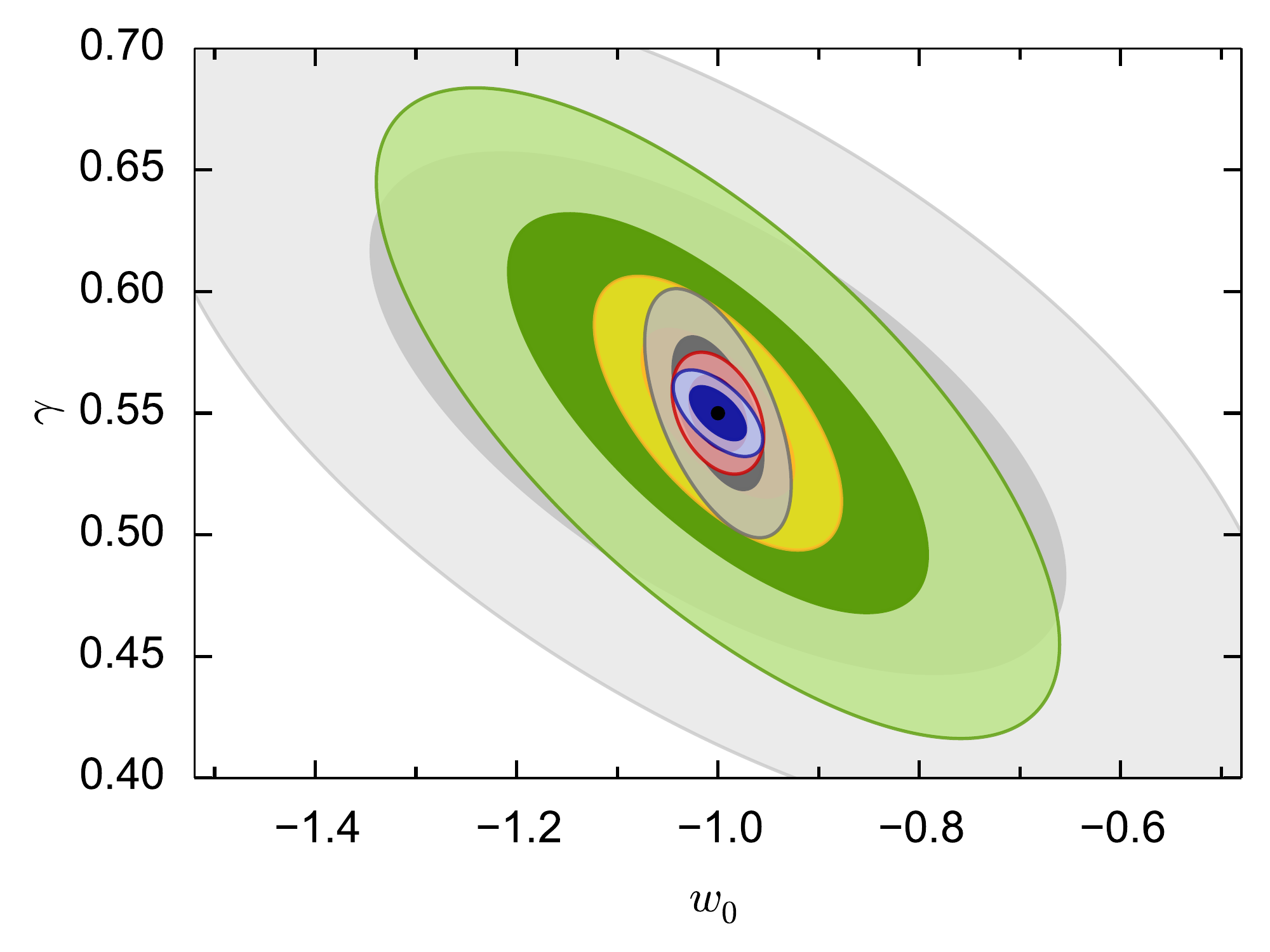}
\caption{Joint constraints on $w_0$, $w_a$, and $\gamma$ for the $\gamma(z) = \gamma_0$ growth parametrisation. All of the ellipses include the BOSS and Planck Fisher matrix priors. All other parameters have been marginalised.}\label{fig:w0}}
\end{figure*}

\paragraph{Intermediate redshifts $(2 \lesssim z \lesssim 3.5)$} The expansion is matter-dominated in this range -- dark energy is a subdominant ($<10\%$) contribution to the net energy density of the Universe, and the growth rate approaches unity in GR. Nevertheless, the high-redshift behaviour of the equation of state can be probed here, $w(a) \to w_0 + w_a$, as can a rescaling of the growth rate, $1 + \eta(a) \to 1 + \eta_0 + \eta_1$. There is already a $\sim$ few percent distance constraint at $z=2.4$ from the BOSS Lyman-$\alpha$ BAO measurement \citep{2013JCAP...04..026S}, and the planned HETDEX survey\footnote{\url{http://hetdex.org/}} should provide similar precision on expansion and growth over most of this range.

\paragraph{Post-reionisation $(3.5 \lesssim z \lesssim 6)$} This is physically similar to the previous redshift range except that the galaxy population is younger. While some MG/dark energy theories may modify the expansion rate at these high redshifts \citep{2013PhRvD..87h3505C, 2013CQGra..30u4003T}, this regime is more likely to be useful as a consistency check on the lower redshift constraints, e.g. as a way of resolving degeneracies between massive neutrinos and modified gravity \citep{2014MNRAS.440...75B, 2014PhRvD..90b3528B, 2015arXiv150705102V}. There are no existing expansion/growth constraints at these redshifts.

{\corr The specifications of the surveys that will cover the various regimes can be found in Sects.~\ref{sec:skaconfig}--\ref{sec:fisherim} (summarised in Table~\ref{tab:skaarrays}), and their redshift coverage is illustrated in Fig.~\ref{fig:surveyz}.}

{\corr Note that forecast constraints on $D_A(z)$, $H(z)$, the dark energy equation of state, and $f\sigma_8$, have been presented before for several of the SKA surveys listed here -- for example, in \cite{Bull:2015nra}, \cite{Raccanelli:2015hsa}, and \cite{2015MNRAS.450.2251Y}. The key difference in this work is that the specifications have been substantially updated following the SKA rebaselining procedure (Sect.~\ref{sec:skaconfig}). Specifically, the previous works used the baseline specifications for MID and LOW, which have seen reductions in collecting area of approximately 30\% and 50\% respectively.}

\subsection{Background expansion}

Fig.~\ref{fig:Hz} shows the forecast constraints on the expansion rate, $H(z)$, and angular diameter distance, $D_{\rm A}(z)$. At low redshifts, an HI galaxy survey with SKA1-MID Band 2 will perform worse than current optical surveys owing to its low survey volume, as was discussed in \cite{2015MNRAS.450.2251Y}. A 25,000 deg$^2$ intensity mapping survey in autocorrelation mode with SKA1-MID Band 2 will provide highly competitive $\sim\!\! 1\%$ constraints out to $z \approx 0.7$ however, assuming that foregrounds and other systematics can be handled without causing too large a loss in effective sensitivity.

Similar precision on $H(z)$ could be achieved out to $z \approx 2$ with an equivalent autocorrelation-mode IM survey on MID Band 1, although this is less competitive because H$\alpha$ surveys like Euclid will provide significantly better ($\sim\!\! 0.4\%$) constraints in the same redshift range in the same timeframe. The MID Band 1 constraints on $D_{\rm A}$ degrade more rapidly with redshift due to the resolution effects discussed in Sect.~\ref{sec:fisherim}, but remain at the few-percent level. Note that the Alternative SKA1-MID configuration would result in a small improvement in $H$ and $D_{\rm A}$ constraints over the Rebaselined specification, although the loss of redshift bins at $z<0.7$ would decrease the MID Band 1 dark energy figure of merit.

More promising is an SKA2 HI galaxy survey, which will measure $H(z)$ with sub-percent precision over 30,000 deg$^2$ from $z\approx0.1-2.0$, reaching the $0.3\%$ level around $z\approx1$ where sensitivity to the dark energy equation of state parameters peaks (according to Fig.~\ref{fig:derivs}). This supports the classification of SKA2 as a key Stage IV dark energy survey by the Dark Energy Task Force \citep[DETF;][]{2006astro.ph..9591A} if the final specifications are comparable to what we assumed here.

At higher redshifts $(z \gtrsim 2)$, IM surveys on SKA1-MID Band 1 yield $1-3\%$ precision on $H(z)$ for the Rebaselined configuration, at least equalling the performance of HETDEX in the same range (albeit with a lower maximum redshift). The $D_{\rm A}(z)$ constraint is worse than for HETDEX however, and the Alternative MID Band 1 configuration would not be able to reach beyond $z=2$ at all. SKA1-LOW is less sensitive, producing only $5-6\%$ constraints on $H(z)$, but covers a wide (and as-yet unprobed) redshift range, so could be useful in constraining MG/DE models that deviate strongly from GR+$\Lambda$CDM only at high redshift. Recall that the performance of IM surveys is contingent on the efficiency of foreground subtraction, calibration, and so on (which are expected to be more difficult at lower frequencies). We have not accounted for these effects in our forecasts, so the figures for MID and LOW should be seen as a `best-case' scenario.

Note that some of the configurations have the same sensitivity at a given redshift, but exhibit small differences in their $H(z)$ constraints (e.g. compare the Updated and Alternative configurations of SKA1-LOW in Fig.~\ref{fig:Hz}). This is caused by correlations between $H(z)$ and $\sigma_{\rm NL}$, the nuisance parameter representing non-linear effects. The SKA1-LOW Alternative configuration constrains $\sigma_{\rm NL}$ better due to its extended redshift coverage, for example, which results in a slight improvement of the marginalised $H(z)$ constraints.

\subsection{Dark energy equation of state}

Fig.~\ref{fig:w0} shows corresponding forecasts for the equation of state parameters $w_0$ and $w_a$, including Planck CMB and BOSS LSS Fisher matrix priors. The curvature, $\Omega_K$, and growth index, $\gamma_0$, have been marginalised over. A summary of the 1D marginal constraints on $w_0$ and $w_a$ is also given in Table \ref{tab:marginals}, assuming different sets of growth parameters.

\begin{figure*}[t]
\centering{
\includegraphics[width=1.03\columnwidth]{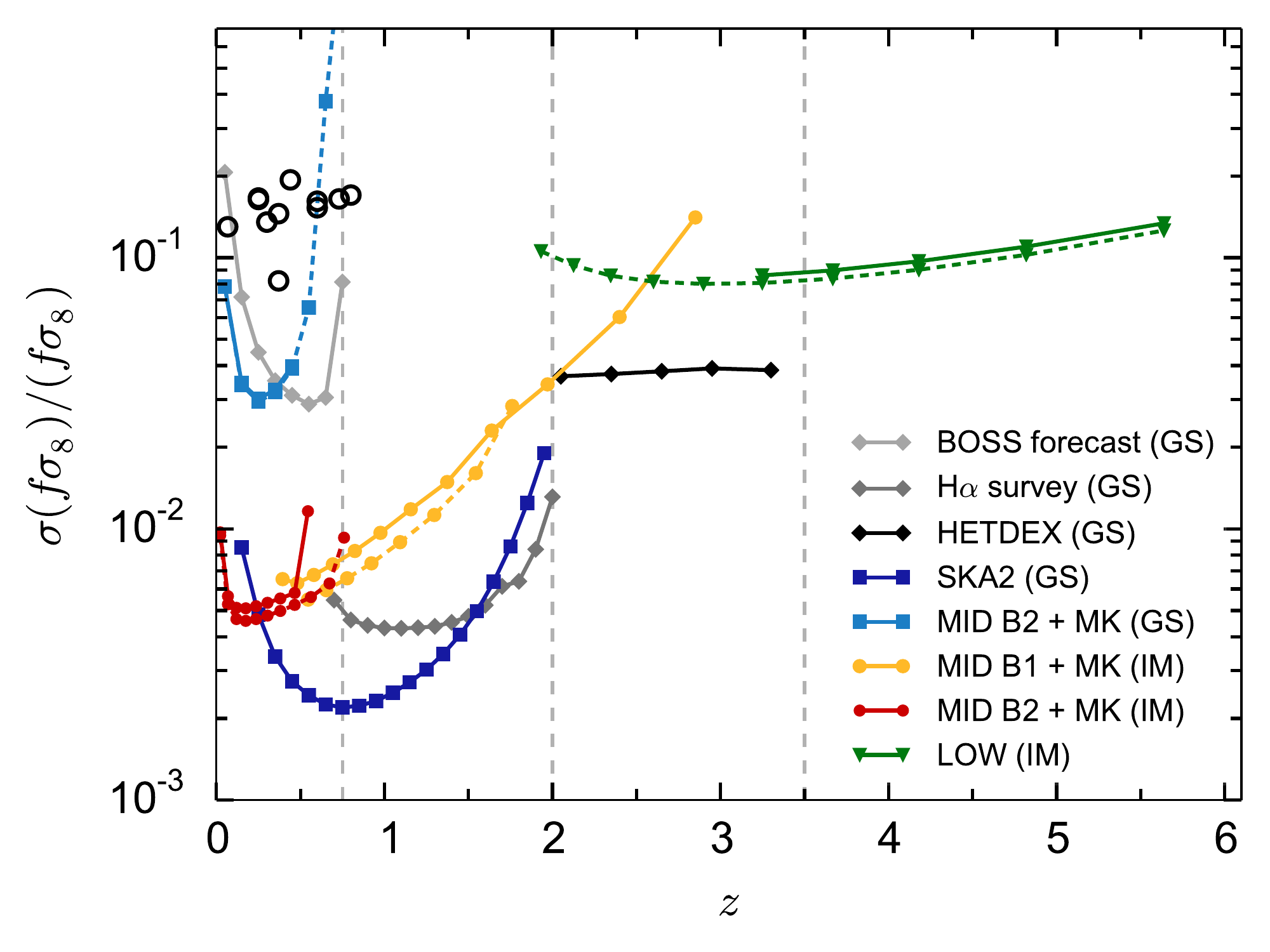}
\includegraphics[width=1.04\columnwidth]{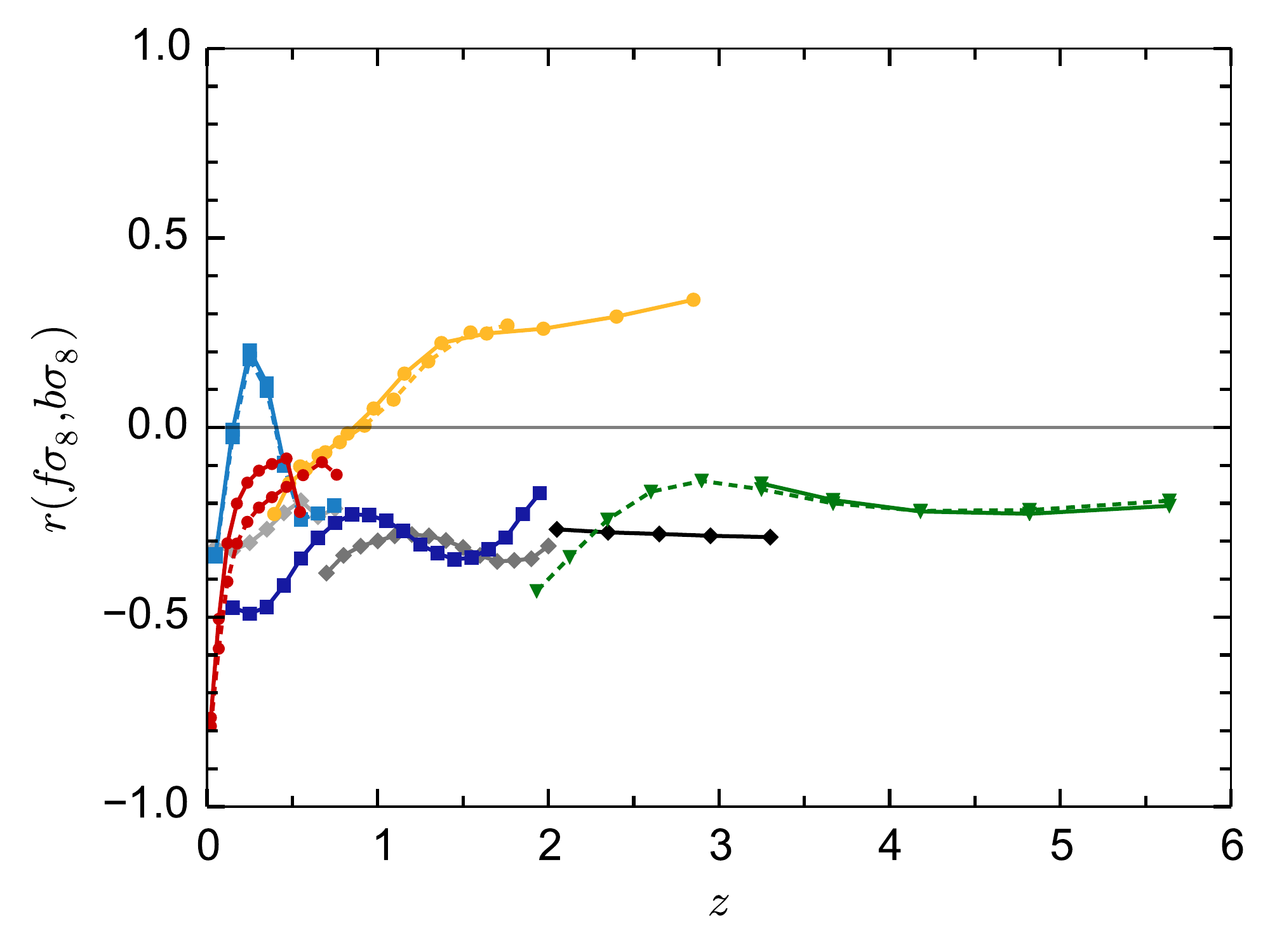}
\caption{{\it Left panel:} Forecast constraints on $f \sigma_8$ for the various SKA galaxy surveys (GS) and intensity mapping surveys (IM). For SKA1-MID surveys, dashed lines show the `Alternative' configuration. Unfilled circles show the errors on existing measurements of $f \sigma_8$ from large scale structure surveys, taken from the compilation in \cite{2013PhRvL.111p1301M}. {\it Right panel:} Correlation coefficient, $r(x,y) = \sigma_{xy} / (\sigma_x \sigma_y)$, between the growth rate ($f\sigma_8$) and bias ($b\sigma_8$) as a function of redshift.}
\label{fig:fs8}
}
\end{figure*}

The sensitivity of SKA1-LOW to $H(z)$ and $D_{\rm A}(z)$ is relatively low, so it has little to add over BOSS in terms of constraints on the equation of state parameters. SKA1-MID Band 1 (Alternative configuration) offers a significant improvement thanks to its stronger (sub-1\%) constraints on the background expansion and more suitable redshift range -- leading to a $5-6\%$ constraint on $w_0$ -- but as anticipated by Fig.~\ref{fig:Hz}, the H$\alpha$ survey outperforms it, reaching $3-4\%$ precision on $w_0$. The difference is not as drastic as one might expect, however, mostly thanks to MID's lower $z_{\rm min}$ compensating for its lower sensitivity at higher $z$. As expected, the SKA2 HI galaxy survey consistently offers the tightest constraints ($\sigma(w_0) \approx 0.02$, $\sigma(w_a) \approx 0.08$) because of its high sensitivity, large survey volume, and essentially ideal redshift range.

Most interesting is the SKA1-MID Band 2 IM survey, which yields exceptionally strong constraints on $(w_0, w_a)$, approaching those of the SKA2 HI galaxy survey, or even surpassing it for some parametrisations. This is mostly due to its ability to break certain degeneracies by reaching the very lowest redshifts. While the forecast errors on $H(z)$ and $D_{\rm A}(z)$ are larger than those of the SKA2 galaxy survey by a factor of $2-3$, the high precision of MID Band 2 at $z \approx 0$ yields tighter constraints on other parameters, notably $H_0$, that $w_0$ and $w_a$ are correlated with. Adding a tight \citep[but achievable;][]{2012arXiv1202.4459S} external prior on $H_0$ at the $\sim\!\!0.5\%$ level breaks the degeneracy for other surveys too, reducing (but not completely removing) SKA1-MID's advantage over the H$\alpha$ galaxy survey for example.

Note that the constraints on $w_0$ and $w_a$ are insensitive to the choice of growth parametrisation. Even the relatively loose constraint on $w_a$ from SKA1-MID Band 1 changes by less than one percentage point across all four growth parametrisations in Table \ref{tab:marginals}.

Now that we have an overview of how well various experiments can measure the dark energy equation of state, it is pertinent to ask how effective these constraints will be in actually testing realistic dark energy and modified gravity models. This is a difficult question to answer however, as the space of plausible models is extremely broad and complex \citep[e.g. see][]{Clifton:2011jh}, and many models allow expansion histories that are arbitrarily close to GR+$\Lambda$CDM. Taking the simplest class of minimally coupled scalar field quintessence models, \cite{Marsh:2014xoa} showed that there is no `target uncertainty' on $w_0$ and $w_a$ beyond which most alternative models could be distinguished from each other or ruled out. There is therefore no particular precision level to aim for -- we can only continue to improve $w(z)$ constraints in the hope that some deviation from GR+$\Lambda$CDM will turn up to give us more clues about what to look for. The role of the SKA will be to substantially improve constraints on $w(z)$ in the key low redshift regime, extend measurements to higher redshifts with LOW, and then to provide the highest precision observations at $z<2$ with an SKA2 HI galaxy survey.

\subsection{Linear growth rate}

Forecasts for the growth rate, $f\sigma_8$, are shown in Fig.~\ref{fig:fs8}. Existing constraints from the literature are also shown, based on the compilation in \cite{2013PhRvL.111p1301M}. Conservatively, the bias has been marginalised as a free parameter in each redshift bin in all of the forecasts, as described in Sect. \ref{sec:params}.

The existing constraints are restricted to the lowest redshifts only, $z \lesssim 0.8$, and are at the $10-20\%$ level. The SKA1-MID galaxy surveys mildly improve on this, reaching the few percent level at the same redshifts. The MID Band 2 IM survey is again much more powerful, yielding $0.5-0.6\%$ constraints at $z \gtrsim 0.2$ that are insensitive to the assumed configuration. A precision of $0.6-3\%$ is achievable with the MID Band 1 IM survey over $0.7 \lesssim z \lesssim 2$, but again this is substantially bettered by the H$\alpha$ survey, which reaches 0.4\% over much of the same redshift range.

The SKA2 galaxy survey again provides the tightest constraints over a wide redshift range, reaching the $\sim0.3\%$ level, although the H$\alpha$ survey can slightly surpass it at $z \approx 2$. At $z > 2$, HETDEX can place $\sim 4\%$ constraints on $f\sigma_8$ out to $z \approx 3.5$, which is roughly a factor of 2 better than the Alternative configuration of SKA1-LOW. Depending on the final band specification, LOW can put $\sim10\%$ constraints on $f\sigma_8$ over the entire redshift range $2 < z < 6$ however, which is beyond the capabilities of any other survey.

As mentioned previously, the bias is an important source of uncertainty in large-scale structure analyses. The right panel of Fig.~\ref{fig:fs8} shows the correlation coefficient between $f\sigma_8$ and the bias, $b\sigma_8$, in each redshift bin. In all cases the correlation is moderate, except for the SKA1-MID Band 2 IM survey at low redshift, where a strong anti-correlation arises. We have already been conservative in marginalising over the bias in each redshift bin, so we do not expect our predictions to strongly depend on the assumed bias model.

As with the equation of state, it is useful to consider how well the $f\sigma_8$ constraints will be able to distinguish between different dark energy and modified gravity models. \cite{2015arXiv150603047P} randomly generated thousands of Horndeski EFT (effective field theory) models subject to the condition that they meet a set of viability criteria, and then calculated the distribution of $f\sigma_8(z)$ functions that the models predict (see their Fig. 4). While their analysis is not entirely general (it depends on a particular parametrisation for the evolution of the EFT coupling functions), it does give some idea of the `typical' range of $f\sigma_8(z)$ for a broad class of MG models, and so we will use their results for illustration.

\begin{table*}[t]
\begin{center}
{\renewcommand{\arraystretch}{1.4}
\begin{tabular}{c|rrrr|rrrr|rrrr|rrrr|}
 \multicolumn{1}{l|}{ } & \multicolumn{4}{c|}{\bf $\bm{\gamma_0}$ free} & \multicolumn{4}{c|}{\bf $\bm{(\gamma_0, \gamma_1)}$ free} & \multicolumn{4}{c|}{\bf $\bm{(\gamma_0, \eta_0)}$ free} & \multicolumn{4}{c|}{\bf $\bm{(\eta_0, \eta_1)}$ free} \\
\hline
 & \multicolumn{16}{c|}{\it incl. Planck} \\
\hline
  & MID1 & MID2 & SKA2 & H$\alpha$ & MID1 & MID2 & SKA2 & H$\alpha$ & MID1 & MID2 & SKA2 & H$\alpha$ & MID1 & MID2 & SKA2 & H$\alpha$ \\
$\bm{h}$ & 0.7 & 0.3 & 0.2 & 0.4 & 0.9 & 0.5 & 0.4 & 0.8 & 0.9 & 0.5 & 0.4 & 0.8 & 0.9 & 0.5 & 0.4 & 0.8 \\
\bm{$w_0}$ & 5.4 & 1.9 & 1.8 & 3.1 & 5.7 & 1.9 & 2.4 & 4.6 & 5.7 & 1.9 & 2.2 & 4.6 & 5.7 & 1.9 & 2.2 & 4.6 \\
$\bm{w_a}$ & 14.3 & 9.2 & 6.9 & 11.0 & 15.2 & 9.4 & 8.4 & 13.9 & 15.3 & 9.2 & 8.4 & 14.1 & 15.3 & 9.3 & 8.4 & 14.0 \\
\hline
$\bm{\gamma_0}$ & 2.5 & 1.1 & 0.7 & 2.4 & 4.2 & 1.2 & 1.0 & 3.4 & 2.8 & 1.7 & 0.9 & 2.4 & --- & --- & --- & --- \\
$\bm{\gamma_1}$ & --- & --- & --- & --- & 10.9 & 6.1 & 4.4 & 10.3 & --- & --- & --- & --- & --- & --- & --- & --- \\
\hline
$\bm{\eta_0}$ & --- & --- & --- & --- & --- & --- & --- & --- & 1.3 & 1.4 & 0.7 & 1.3 & 2.3 & 1.2 & 0.8 & 2.4 \\
$\bm{\eta_1}$ & --- & --- & --- & --- & --- & --- & --- & --- & --- & --- & --- & --- & 4.1 & 3.3 & 1.4 & 3.2 \\
\hline
 & \multicolumn{16}{c|}{\it incl. Planck + BOSS} \\
\hline
  & MID1 & MID2 & SKA2 & H$\alpha$ & MID1 & MID2 & SKA2 & H$\alpha$ & MID1 & MID2 & SKA2 & H$\alpha$ & MID1 & MID2 & SKA2 & H$\alpha$ \\
$\bm{h}$ & 0.6 & 0.3 & 0.2 & 0.4 & 0.8 & 0.4 & 0.4 & 0.7 & 0.8 & 0.4 & 0.4 & 0.7 & 0.8 & 0.4 & 0.4 & 0.7 \\
\bm{$w_0}$ & 5.0 & 1.8 & 1.8 & 2.9 & 5.2 & 1.8 & 2.3 & 4.2 & 5.1 & 1.9 & 2.2 & 4.2 & 5.1 & 1.9 & 2.2 & 4.2 \\
$\bm{w_a}$ & 13.6 & 9.1 & 6.8 & 10.1 & 14.4 & 9.2 & 8.2 & 12.3 & 14.6 & 9.1 & 8.3 & 13.0 & 14.5 & 9.1 & 8.3 & 12.9 \\
\hline
$\bm{\gamma_0}$ & 2.3 & 1.0 & 0.7 & 2.1 & 3.5 & 1.2 & 1.0 & 2.7 & 2.6 & 1.7 & 0.8 & 2.1 & --- & --- & --- & --- \\
$\bm{\gamma_1}$ & --- & --- & --- & --- & 9.4 & 6.0 & 4.3 & 9.3 & --- & --- & --- & --- & --- & --- & --- & --- \\
\hline
$\bm{\eta_0}$ & --- & --- & --- & --- & --- & --- & --- & --- & 1.2 & 1.4 & 0.7 & 1.2 & 2.1 & 1.1 & 0.8 & 2.2 \\
$\bm{\eta_1}$ & --- & --- & --- & --- & --- & --- & --- & --- & --- & --- & --- & --- & 3.8 & 3.2 & 1.4 & 2.9 \\
\hline
\end{tabular} }
\end{center}
\caption{Marginal errors ($1\sigma \times 100$) on cosmological parameters for various parametrisations, calculated for four experiments: SKA1-MID B1 Alt. (IM); SKA1-MID B2 Alt. (IM); SKA2 (GS); and a H$\alpha$ survey (GS).}
\label{tab:marginals}
\end{table*}

According to \cite{2015arXiv150603047P}, for $z \approx 1-2$ a precision of $\sim\!\!10\%$ on $f\sigma_8$ would be required to start ruling out a significant number of models from the {\corr general} Horndeski 4/5 class\footnote{{\corr This class includes all Horndeski scalar field theories, including Brans-Dicke, $f(R)$ gravity, quintessence, K-essence, Galileons and others.}} (see that paper for a definition). This target is well within the reach of the SKA1-MID Band 1 IM survey, and H$\alpha$ and SKA2 galaxy surveys. The more restrictive Brans-Dicke subclass predicts much less variation in $f\sigma_8$ in this range though, requiring $\sim\! 0.5\%$ precision to make any headway. The picture at low redshift is more encouraging however, with a $\sim\!\! 5\%$ constraint at $z=0.4$ being sufficient to rule out a sizeable fraction of models from both the Brans-Dicke and Horndeski 4/5 classes. The models predict a large spread of values of $f\sigma_8$ at $z\lesssim 0.2$, so the SKA1-MID Band 2 IM survey could be particularly powerful in constraining these theories.

\begin{figure}[t]
\hspace{-2em}\includegraphics[width=1.04\columnwidth]{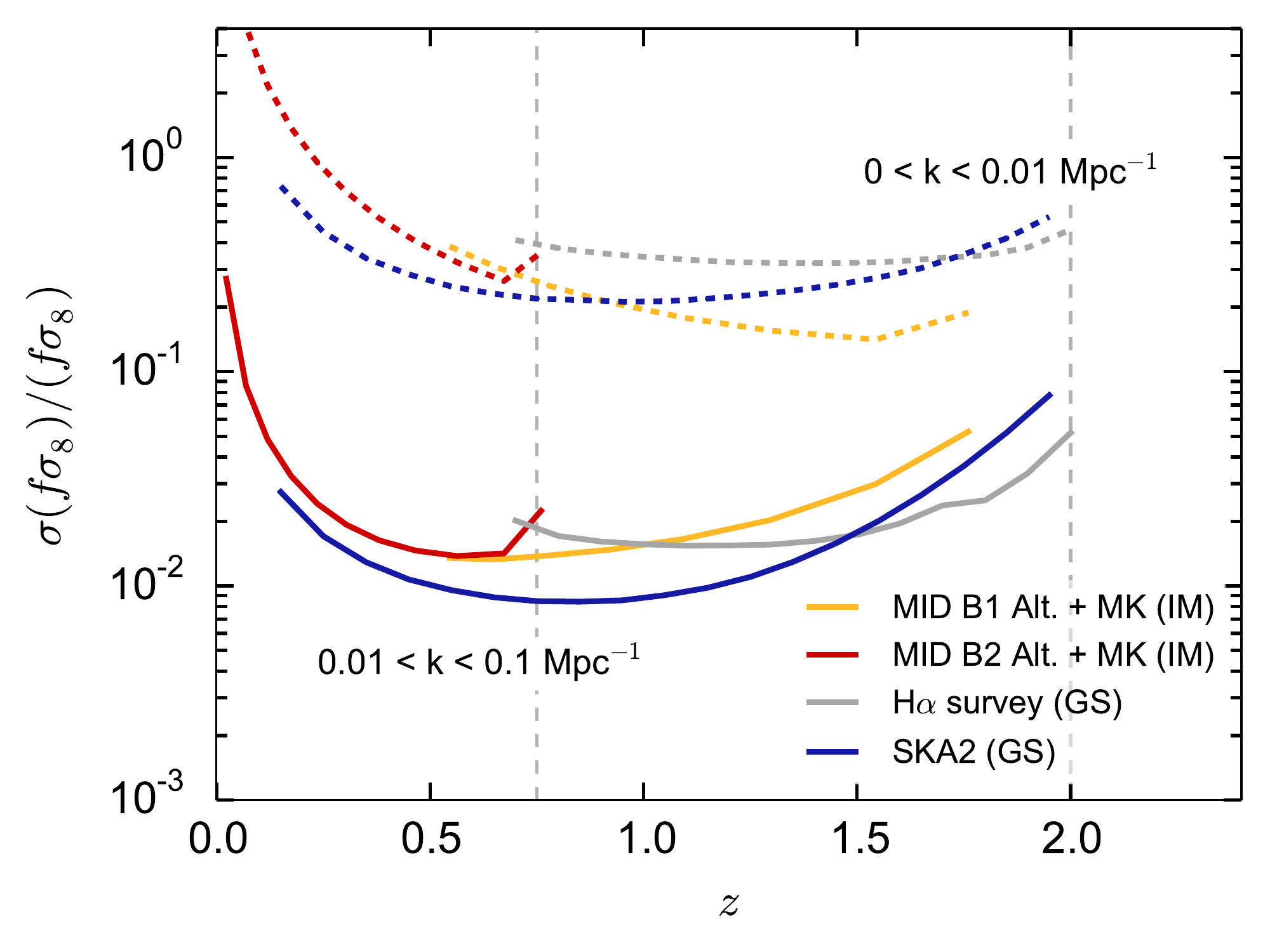}
\caption{Constraints on $f\sigma_8$ for several surveys, binned by scale.}
\label{fig:fs8-scaledep}
\end{figure}

Finally, we present constraints on the growth rate as a function of scale in Fig.~\ref{fig:fs8-scaledep}. The bias, $b\sigma_8$, {\corr was also binned} in scale and marginalised over. On linear, quasi-static scales ($0.01 < k < 0.1$ Mpc$^{-1}$), the constraints from all four surveys that we consider (SKA1-MID Band 1 and 2 IM surveys, and the H$\alpha$ and SKA2 galaxy surveys) are broadly similar -- all at the $1-2\%$ level {\corr for $z \approx 1$}. This is in contrast to the scale-independent constraints, where the differences between surveys are more pronounced. This is because of the better sensitivity of the galaxy surveys to smaller scales ($k \ge 0.1$ Mpc$^{-1}$; not shown), where there are more Fourier modes -- autocorrelation-mode IM surveys are less sensitive to small scales due to their limited angular resolution. On large scales ($k < 0.01$ Mpc$^{-1}$) the constraints are worse by an order of magnitude ($\sim 30\%$, instead of $\sim 1\%$), but otherwise follow a similar pattern. Note that MID Band 1 does become more sensitive than the H$\alpha$ survey at $z\approx2$ on these scales due to its larger survey area, although the foreground removal required by IM experiments is least efficient at small $k$, potentially removing this advantage \citep[galaxy surveys also suffer from large-scale systematics though; see][]{2015MNRAS.451..539G}.

\begin{table}[t]
\begin{center}
{\renewcommand{\arraystretch}{1.6}
\begin{tabular}{|l|r|r|r|r|r|r|}
\multicolumn{1}{r|}{Fiducial $k_\mu = $} & \multicolumn{2}{c|}{$0.005$ Mpc$^{-1}$} & \multicolumn{2}{c|}{$0.01$ Mpc$^{-1}$} & \multicolumn{2}{c|}{$0.02$ Mpc$^{-1}$} \\
\hline
{\bf Survey} & \multicolumn{1}{c|}{$A_\mu$} & \multicolumn{1}{c|}{$\log k_\mu$} & \multicolumn{1}{c|}{$A_\mu$} & \multicolumn{1}{c|}{$\log k_\mu$} & \multicolumn{1}{c|}{$A_\mu$} & \multicolumn{1}{c|}{$\log k_\mu$} \\
\hline
LOW Alt.          & 0.430 & 89.2~~ & 0.430 & 20.4~~ & 0.430 & 6.55 \\
MID B1 Rebase.    & 0.039 & 0.64 & 0.028 & 0.39 & 0.003 & 0.03 \\
MID B2 Rebase.    & 0.094 & 2.27 & 0.095 & 1.57 & 0.048 & 0.71 \\
MID B1 Alt.       & 0.045 & 0.84 & 0.045 & 0.70 & 0.012 & 0.16 \\
MID B2 Alt.       & 0.055 & 1.30 & 0.055 & 0.89 & 0.020 & 0.26 \\
SKA2              & 0.017 & 0.43 & 0.017 & 0.28 & 0.005 & 0.06 \\
H$\alpha$ survey  & 0.038 & 0.84 & 0.038 & 0.63 & 0.014 & 0.20 \\
\hline
\end{tabular} }
\end{center}
\vspace{-1em}\caption{$1\sigma$ errors on $A_\mu$ and $\log_{10} k_\mu$ for a fiducial amplitude $A_\mu = 0.01$ and several fiducial values of $k_\mu$. {\corr Only the Planck priors} are included.}\vspace{-1em}
\label{tab:scaledep}
\end{table}

\subsection{Growth parametrisations}

Table \ref{tab:marginals} shows marginal constraints on several parameters, for four different growth parametrisations. These can be compared with previous works \citep[e.g.][]{PhysRevD.77.083508}.

As with the equation of state parameters, an SKA1-MID Band 2 IM survey is capable of putting remarkably stringent constraints on the growth parameters, rivalling the considerably more precise SKA2 HI galaxy survey for all four parametrisations. This is again an artefact of its high precision at the lowest redshifts, which strongly breaks degeneracies with parameters like $H_0$. Adding the BOSS prior reduces this advantage slightly, bringing the H$\alpha$ survey results to within a factor of $\sim2$ of the MID Band 2 constraints.

Forecasts for the `theoretically-motivated' parameters $(A_\mu, k_\mu)$ are presented in Table \ref{tab:scaledep} for a small fiducial deviation amplitude $A_\mu = 10^{-2}$ (i.e. a 1\% correction to the Poisson equation at $z=0$), and several different scale parameters, $k_\mu$. The Planck prior has been included (but not the BOSS one), and $w_0$, $w_a$, and $\Omega_K$ have been marginalised as usual. The two MG parameters are correlated; fixing $k_\mu$ typically improves the forecast errors on $A_\mu$ by a factor of a few.

None of the surveys will be capable of detecting a deviation from GR at this level on their own unless $k_\mu$ is quite large, suggesting that a multi-tracer approach \citep{2009JCAP...10..007M} may be necessary to improve precision. Even the SKA2 constraint of $\sigma(A_\mu) = 0.017$ for $k_\mu = 0.01$ Mpc$^{-1}$ falls shy of a $1\sigma$ detection. Fixing $w_0$, $w_a$, and $\Omega_K$ also makes little difference. This is because the error on $A_\mu$ is mostly driven by the bias uncertainty -- fixing the bias improves the SKA2 + Planck constraint by roughly a factor of two. For comparison, a full-sky cosmic variance-limited galaxy survey from $z=0-3$ (with $b(z) = \sqrt{1+z}$) would achieve $\sigma(A_\mu) = 0.004$ with or without the bias marginalised. We conclude that, while a multi-tracer analysis should be helpful, the more immediate limitation is probably the bias rather than cosmic variance.

In \cite{2013MNRAS.429.2249S}, the RSD-only constraint on their $\mu_0$ parameter (similar to our $A_\mu$) is based on five $\sim\!20\%$ measurements of $f\sigma_8$ at $z < 0.8$ from WiggleZ and 6dFGS. They find $\sigma_{\mu0} = 0.25$ for a $\Lambda$CDM background. \cite{2015arXiv150201590P} improve this to $\sigma_{\mu 0} = ^{+0.12}_{-0.21}$ by adding more datasets (see their Table 6, for the parameter $\mu_0 - 1$). For comparison, our BOSS forecast \citep[based on the BOSS specification in][]{Font-Ribera:2013rwa} predicts $\sigma(A_\mu) = 0.04$ (with fixed $k_\mu = 0.01$ Mpc$^{-1}$) from seven redshift bins with $\sim 4\%$ measurements of $f\sigma_8$. The actual BOSS dataset yielded RSD measurements in only a couple of redshift bins however, with $\sim\!\!10\%$ errors on $f\sigma_8$ \citep{2012MNRAS.424.2339T, Samushia:2013yga}. This gives some idea of the level of optimism of our forecasts: real analyses can lose a great deal of sensitivity compared to the `ideal' Fisher approach, e.g. due to cuts and additional modelling uncertainties introduced as a by-product of correcting for systematic effects. While one can hope that analysis methods become less `lossy' in future, it is likely that systematics will in fact get more complex and numerous as raw observational sensitivity improves. Our forecast precision levels should therefore be considered achievable only up to a factor of a few.

Finally, the scale dependence can be constrained to within an order of magnitude by both SKA2 and the H$\alpha$ survey, regardless of the fiducial scale $k_\mu$. The Band 1 SKA1-MID IM survey is better at measuring this parameter than Band 2 because of its wider (and higher) redshift range, meaning that it can access a larger volume and thus sample large scales better. The constraints degrade significantly for the smallest fiducial $k_\mu$, as the scale-dependent modification is shifted to larger scales where fewer modes can be measured. It was argued in \cite{2015arXiv150600641B} that a much larger scale of $k_\mu\! \sim\! \mathcal{H} \approx 2 \times 10^{-4}$~Mpc$^{-1}$ is the most natural choice in the absence of a specific MG theory, so the fiducial values we have chosen here are not particularly well-motivated. A proper treatment of scale-dependent MG effects on ultra-large scales requires the relaxation of several assumptions we have made in our forecasting formalism, and so we will leave this to future work.

\section{Discussion} \label{sec:discussion}

Forthcoming large-scale structure surveys will greatly extend the range of scales and redshifts over which General Relativity can be tested. In this paper, we examined how surveys of neutral hydrogen with the Square Kilometre Array (SKA) -- using either spectroscopic galaxy redshifts or 21cm intensity maps -- can be used to probe dark energy and modified gravity theories from $z=0$ to $z \approx 6$.

Using galaxy clustering and redshift-space distortions as our observables, we performed Fisher forecasts for the expansion rate, angular diameter distance, and growth rate as a function of redshift and scale, which we then mapped onto several phenomenological parametrisations of modified gravity and dark energy theories. Two possible designs -- `Rebaselined' and `Alternative' -- were considered to account for uncertainty in the SKA Phase 1 specifications.

By way of an `executive summary', we will now discuss the implications of our forecasts for each of the proposed SKA cosmology surveys in turn:
\begin{itemize}[leftmargin=*]
 \item We confirm the finding of \cite{Bull:2015nra, Raccanelli:2015hsa} that an {\bf SKA1 HI galaxy survey} will not significantly improve existing low-$z$ constraints from optical surveys on its own -- its maximum redshift and mean number density are simply too low. It may be useful if cross-correlated with other surveys though: a multi-tracer analysis can lead to large gains in precision of the growth rate measurement if the ratio of the biases of the cross-correlated tracer populations is large \citep{2009JCAP...10..007M}. The low bias of the HI galaxies ($b \approx 0.7-0.9$) may therefore be a useful resource. Multi-tracer analyses also require high number densities (achievable for SKA1-MID only at $z \lesssim 0.3$) and substantial survey area overlap (possible for most future LSS and CMB surveys thanks to the SKA's location in the southern hemisphere).

 \item A more promising prospect at low redshift is an {\bf SKA1-MID Band 2 IM} (intensity mapping) survey, which could map out the large scale HI distribution over 25,000 deg$^2$ out to $z\approx 0.6-0.8$ with high signal-to-noise in 10,000 hours. This would require the MID dish array to operate in autocorrelation mode, which has yet to be tested on a large radio array, and which will require a considerable amount of further development work to deliver sufficiently good calibration and noise properties \citep{Bull:2014rha}. Though more risky from this perspective, a MID Band 2 IM survey offers the best prospect of delivering cosmological constraints with Phase 1 of the SKA that are competitive with contemporary experiments like Euclid; our forecasts show that it outperforms the H$\alpha$ survey for almost all of the dark energy/modified gravity parameters we considered (see Table \ref{tab:marginals}). This is mainly due to its high precision at $z \approx 0$, which helps to break degeneracies with parameters like $H_0$ -- suggesting that further improvement of low-$z$ constraints is very useful for testing alternative theories in general.
 
 \item An {\bf SKA1-MID Band 1 IM survey} appears to be less competitive, reporting constraints on the growth and expansion rates that are a factor of a few worse than a H$\alpha$ spectroscopic survey over the same redshift range. Still, the higher $z_{\rm max}$, large survey area, and therefore substantially larger survey volume of the MID Band 1 survey may make it an important part of a future `multi-tracer' strategy to detect novel relativistic effects or the signatures of primordial non-Gaussianity in the clustering of matter on ultra-large scales \citep{Alonso:2015sfa, 2015arXiv150704605F}, or scale-dependent modifications to GR \citep{2015arXiv150600641B}.


 \item An interferometric {\bf IM survey with SKA1-LOW} would do little to improve constraints on the modified gravity parametrisations that we have adopted, mostly because it only covers $z \gtrsim 2$. It will also produce less competitive growth and expansion constraints than HETDEX in the $2 \lesssim z \lesssim 3.5$ range. Its main use would therefore be to extend measurements of $f\sigma_8(z)$ and $H(z)$ to $3.5 \lesssim z \lesssim 6$, which may be useful for constraining theories with early dark energy evolution, or coupled dark energy/dark matter.

 \item While concrete specifications are yet to be provided, an {\bf SKA Phase 2 HI galaxy survey} looks set to become the `best in class' Stage IV spectroscopic galaxy survey \citep{2006astro.ph..9591A}, thanks to its high predicted number density from $z\approx 0$ to $z\approx 1.6$. Due to its broader redshift range and wide (30,000 deg$^2$) survey area, the SKA2 survey outperforms the H$\alpha$ survey (and all other surveys that we considered) in practically every measure, but as shown in Figs.~\ref{fig:Hz} and \ref{fig:fs8} the improvement is not particularly dramatic -- gaining at most a factor of two over the H$\alpha$ survey forecasts for $H(z)$, $D_{\rm A}(z)$, and $f\sigma_8(z)$ where their redshift coverage overlaps. This does translate to substantially better constraints on the dark energy and modified gravity parameters however (Table \ref{tab:marginals}), typically by a factor of $2-3$, although more dramatic gains are likely to require a multi-tracer approach.
\end{itemize}

Importantly, our forecasts establish that the cosmological survey performance of SKA1 should not be strongly degraded following the `rebaselining' described in \cite{SKArebaselining}. The differences between the results for the Rebaselined and Alternative configurations in Figs.~\ref{fig:Hz} and \ref{fig:fs8} are also small -- although the consequences of altering the frequency bands are potentially more significant, as the redshift range of the surveys is clearly an important factor in the constraints on the modified gravity/dark energy parameters.

In all of our forecasts, we assumed a total survey time of 10,000 hours (approximately 14 months of usable on-sky time), with survey areas chosen to give more or less optimal constraints on the dark energy equation of state (see \cite{Bull:2014rha} and \cite{2015MNRAS.450.2251Y} for optimisations of the IM and HI galaxy surveys respectively). This is very large allotment of time for a general-purpose radio observatory such as the SKA, as many other science goals besides cosmology will be competing for the available observing time. While it is likely that a substantial fraction of the first five years of operation of SKA1 will be dedicated to large survey programmes, it may be necessary to perform the bigger surveys `commensally' (i.e. at the same time). This will inevitably introduce some tension into the choice of survey parameters, with some science goals likely preferring shallow, wide surveys, and others requesting deeper, narrower surveys. Reduced time allocations will also impact on performance. To give some idea of how different survey design assumptions would affect our results, and to inform commensal survey designs, we have presented some measures of survey performance as a function of allotted time and area for SKA1 in Appendix \ref{app:survey}.

Finally, note that in this paper we considered only a limited subset of possible cosmological tests of GR and dark energy: measurements of the background expansion, using distance indicators like BAOs and the broadband shape of the power spectrum; and measurements of the growth rate of perturbations on linear, quasi-static scales, using redshift-space distortions. There are a number of other promising tests that the SKA may be able to perform however, such as: low-redshift Tully-Fisher peculiar velocity surveys with HI galaxies \citep{2014MNRAS.445.4267K}; radio weak lensing\footnote{Weak lensing is particularly complementary to RSDs, as it provides orthogonal constraints in the $(\mu, \Sigma)$ plane, where $\Sigma$ is the modification to the lensing potential, $\Phi + \Psi$ \citep{2013PhRvD..87b3501A, 2013MNRAS.429.2249S, 2015PhRvD..91h3504L}.} surveys \citep{Brown:2015ucq, 2015arXiv150706639H, 2015arXiv150903286P}; observations of relativistic effects on ultra-large scales \citep{Lombriser2013, Alonso:2015uua, Alonso:2015sfa, 2015arXiv150600641B, 2015arXiv150704605F}; `real-time' redshift drift measurements \citep{2015aska.confE..27K}; and strong field and gravitational wave tests using pulsars \citep{2015aska.confE..42S}.

\vspace{-1em}\acknowledgements {\it Acknowledgements ---} I am grateful to T. Baker and M.~G. Santos for extensive discussions, and acknowledge Y. Akrami, S. Camera, P.~G. Ferreira, A. Raccanelli, J.-P. Uzan, M. Viel, F. Villaescusa-Navarro, and K. Zarb-Adami for useful comments and suggestions. PB is supported by European Research Council grant StG2010-257080. The code and Fisher matrices used in this paper are available from \url{http://philbull.com/mg_ska/}.

\appendix
\section{Dependence on survey parameters} \label{app:survey}

\begin{figure*}[t]
\includegraphics[width=0.49\columnwidth]{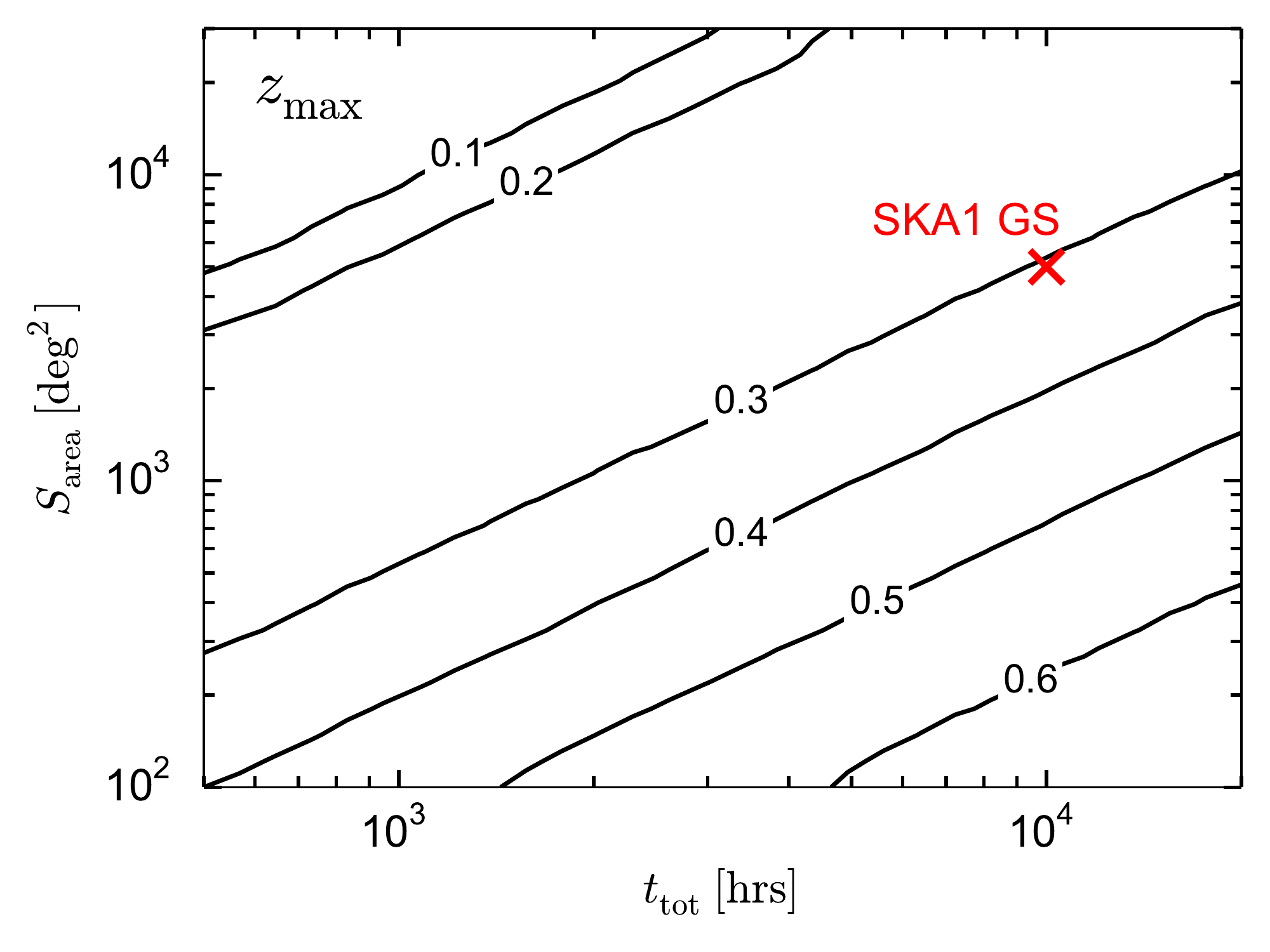}
\includegraphics[width=0.49\columnwidth]{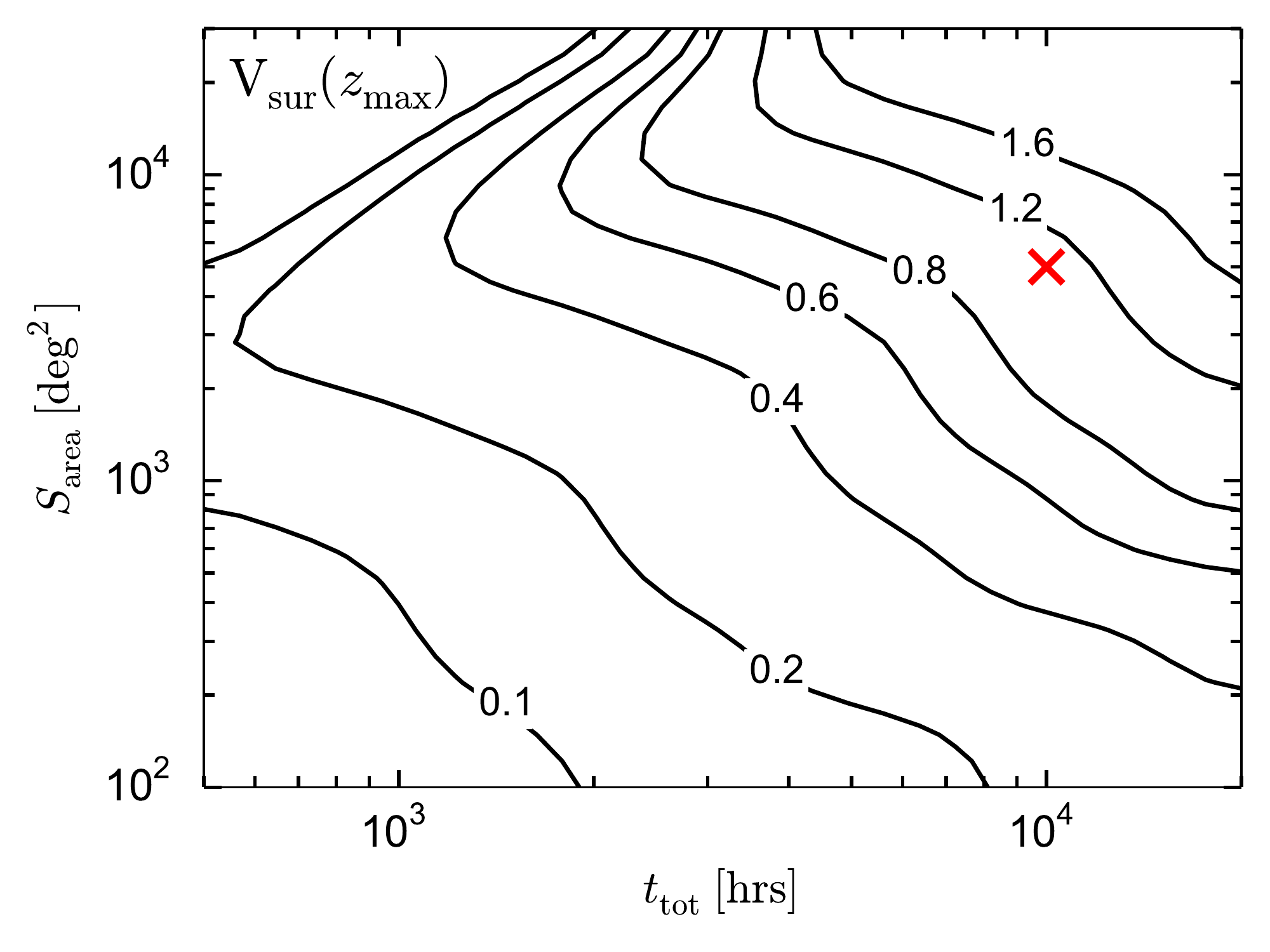}
\caption{The maximum survey redshift and corresponding comoving survey volume, $V_{\rm sur}(0 \!\le\! z \!\le\! z_{\rm max})$, in Gpc$^3$, as a function of survey time and survey area, for an SKA1-MID Band 2 HI galaxy survey in the `Alternative' configuration. The fiducial survey time/area assumed in this study is shown as a red cross. We have assumed that the minimum frequency, $\nu_{\rm min}$, can be adjusted without changing other specifications, and use a detection threshold of $5\sigma$.} 
\label{fig:ska1gs}
\end{figure*}

\begin{figure}[t]
\hspace{1em}\includegraphics[width=0.49\columnwidth]{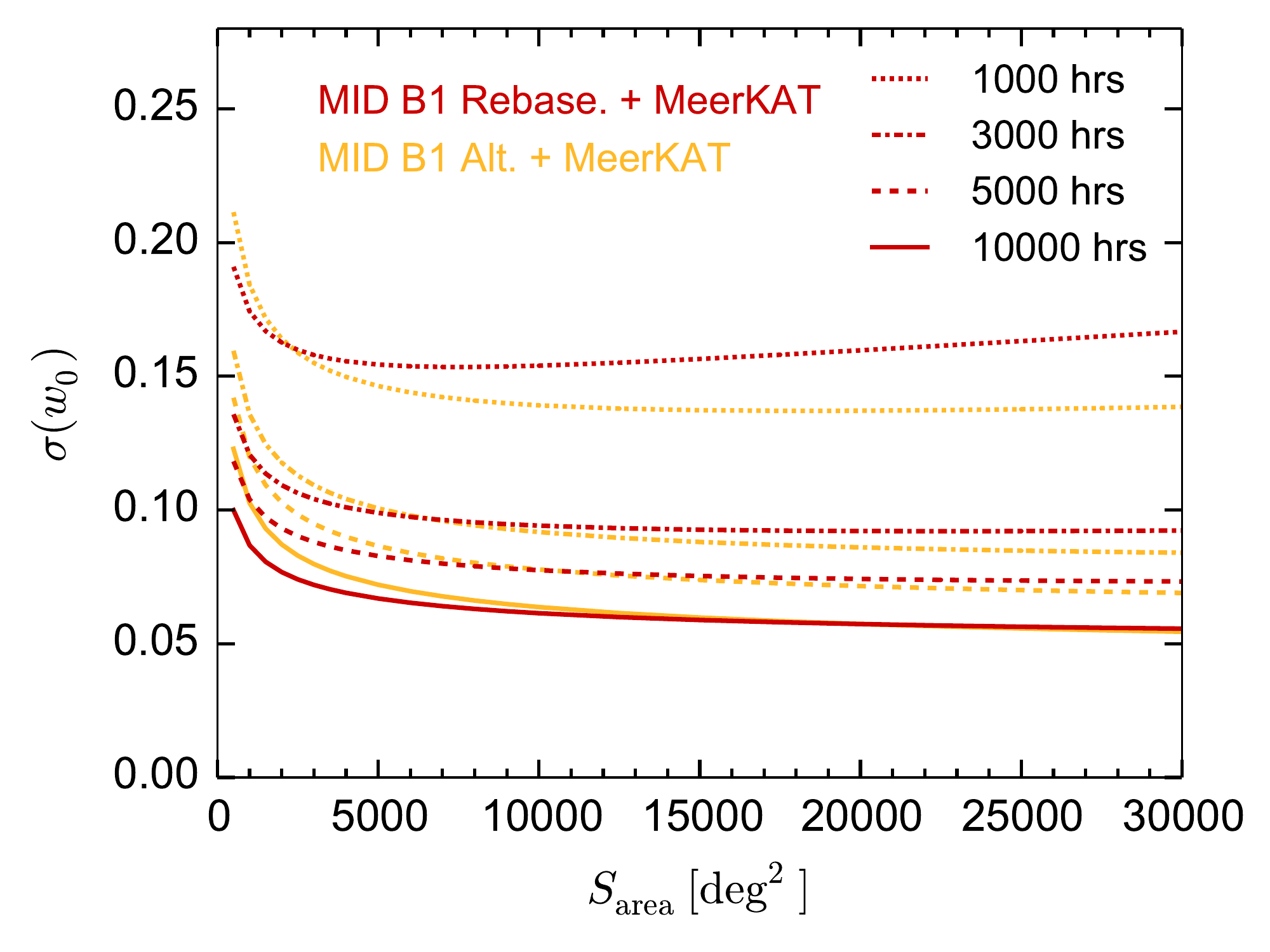}
\includegraphics[width=0.49\columnwidth]{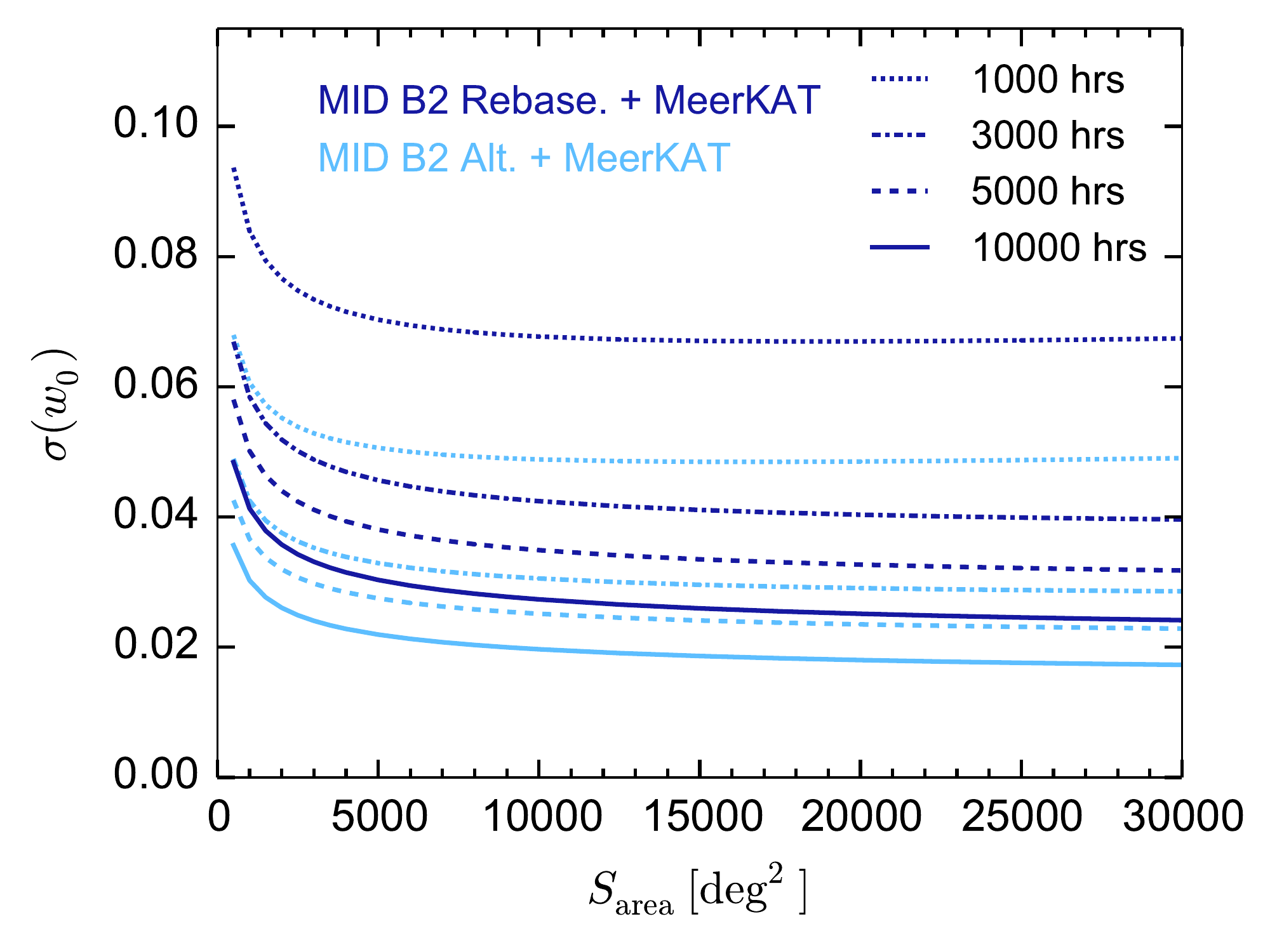}
\caption{Marginal $1\sigma$ error on $w_0$ for several SKA1-MID IM surveys, as a function of survey area and survey time. The Planck prior is included (but not BOSS), and $\gamma_0$ is the only free modified gravity parameter.} 
\label{fig:ska1midw0}
\end{figure}

In this appendix we briefly discuss how the performance of the SKA1 surveys depends on area, $S_{\rm area}$, and survey time, $t_{\rm tot}$.

First we consider the SKA1 HI galaxy survey. Fig.~\ref{fig:ska1gs} shows the maximum redshift, $z_{\rm max}$, at which the non-linear scale is sample variance-limited (i.e. where the signal to noise drops to unity at $k_{\rm NL}(z)$, so that $n(z_{\rm max})\, b^2(z_{\rm max}) P(k_{\rm NL}, z_{\rm max}) \!=\! 1$) as a function of survey time and area. Beyond this redshift, shot noise begins to dominate on progressively larger linear scales. The comoving volume corresponding to this redshift is also shown in the right panel of Fig.~\ref{fig:ska1gs}. Note that $z_{\rm max} \to 0$ in the low $t_{\rm tot}$, high $S_{\rm area}$ region.

For a given observing time, one typically wants to maximise the survey volume over a given redshift range, in order to reduce sample variance. While larger survey areas give higher volumes for a given $t_{\rm tot}$ for SKA1, the maximum redshift can be relatively small, which is problematic if the aim is to measure $H$, $D_{\rm A}$, and $f \sigma_8$ as a function of redshift. Our fiducial choice of 5,000 deg$^2$ attempts to strike a balance, obtaining a relatively large volume while maintaining a reasonable $z_{\rm max}$ of 0.3.

Fig.~\ref{fig:ska1midw0} shows the forecast marginal $1\sigma$ error on $w_0$ for the SKA1-MID intensity mapping surveys, as a function of area for several survey times. While only $w_0$ is plotted here, qualitatively similar results are obtained for other parameters (e.g. $\gamma_0$). Apart from for the shortest survey time, $t_{\rm tot} =$ 1,000 hrs, there is no clear optimum survey area for any of the configurations -- the constraint on $w_0$ improves very slowly with survey area for all $S_{\rm area} \gtrsim 5,000$ deg$^2$. The dependence on survey time is also relatively moderate once a certain threshold is reached; for large survey areas, $\sigma(w_0)$ degrades by a factor of less than 2 for both MID B1 Alt. and MID B2 Alt. when going from $t_{\rm tot} = 10,000$ hrs to $3,000$ hrs.

\section{HI galaxy survey number densities} \label{app:dndz}

In this appendix we present fits to the number density and bias functions derived for the SKA HI galaxy surveys, which are based on the calculations in \cite{2015MNRAS.450.2251Y}. We use the following fitting functions:
\bea
\frac{dN}{dz} &=& 10^{c_1} z^{c_2} \exp({-c_3 z}) ~~~ [{\rm deg}^{-2}]; ~~~~ b(z) = c_4 \exp({c_5 z}).
\eea
The fits were performed using a least-squares procedure, weighted by $\sqrt{dN/dz}$. This produced good fits at redshifts with sizeable number densities, but the fitting functions are poor approximations when $dN/dz$ is small. The results are shown in Table \ref{tab:dndzfits}.

\begin{table}[h]
\begin{center}
{\renewcommand{\arraystretch}{1.6}
\begin{tabular}{|l|r|rrr|rr|r|r|}
\hline
{\bf Survey} & \multicolumn{1}{c|}{\bf Thres.} & \multicolumn{1}{c}{$c_1$} & \multicolumn{1}{c}{$c_2$} & \multicolumn{1}{c|}{$c_3$} & \multicolumn{1}{c}{$c_4$} & \multicolumn{1}{c|}{$c_5$} & \multicolumn{1}{c|}{$z_{\rm max}$} & \multicolumn{1}{c|}{$N_{\rm gal} / 10^6$} \\
\hline
\multirow{2}{*}{SKA1 B2 Rebase. + MeerKAT} & $5\sigma$~~ & 5.450 & 1.310 & 14.394 & 0.616 & 1.017 & 0.391 & 3.49~~~ \\
 & $8\sigma$~~ & 4.939 & 1.027 & 14.125 & 0.913 & -0.153 & 0.329 & 2.04~~~ \\
\hline
\multirow{2}{*}{SKA1 B2 Alt. + MeerKAT} & $5\sigma$~~ & 5.431 & 1.297 & 14.260 & 0.616 & 1.018 & 0.391 & 3.54~~~ \\
 & $8\sigma$~~ & 4.935 & 1.024 & 14.089 & 0.907 & -0.089 & 0.329 & 2.05~~~ \\
\hline
SKA2 & $10\sigma$~~ & 6.319 & 1.736 & 5.424 & 0.554 & 0.783 & 1.084 & 950~~~~~~~~ \\
\hline
\end{tabular} }
\end{center}
\vspace{-1em}\caption{Fitting coefficients for $dN/dz$ and $b(z)$ for the SKA HI galaxy surveys, for various detection thresholds (assuming $S_{\rm area} =$ 5,000 deg$^2$ and 30,000 deg$^2$ for SKA1 and 2 respectively). The maximum redshift at which the non-linear scale is sample variance-limited (see Appendix \ref{app:survey}) and total number of galaxies integrated over the full band are also shown.}
\label{tab:dndzfits}
\end{table}

\bibliographystyle{hapj}
\bibliography{BibliographyMG}

\end{document}